\documentclass[aps,amsmath,amssymb, prc,twocolumn,superscriptaddress]{revtex4-1}

\usepackage{graphicx}
\usepackage{dcolumn}
\usepackage{bm}
\usepackage[utf8]{inputenc}
\usepackage[ngerman,english]{babel}
\usepackage{float}
\usepackage{multirow}
\usepackage{longtable}
\usepackage{dcolumn}
\newcolumntype{d}{D{.}{.}{-1}}
\usepackage{setspace}
\usepackage{threeparttable}
\usepackage{tabularx}
\usepackage[symbol*]{footmisc}
\DefineFNsymbolsTM{myfnsymbols}{
  \textasteriskcentered *
  \textdagger    \dagger
  \textdaggerdbl \ddagger
  \textsection   \mathsection
  \textbardbl    \|%
  \textparagraph \mathparagraph
}%
\setfnsymbol{myfnsymbols}

\usepackage{siunitx}

\begin{document}


\title{Shape coexisistence and collective low-spin states in $^{112,114}$Sn studied with the $(p,p'\gamma)$ DSA coincidence technique}


\author{M. Spieker}
\email[]{spieker@nscl.msu.edu; present address: NSCL, 640 South Shaw Lane, East Lansing, MI 48824, USA}
\affiliation{Institut f\"{u}r Kernphysik, Universit\"{a}t zu K\"{o}ln, Z\"{u}lpicher Straße 77, D-50937 K\"{o}ln, Germany}

\author{P. Petkov}
\affiliation{Institut f\"{u}r Kernphysik, Universit\"{a}t zu K\"{o}ln, Z\"{u}lpicher Straße 77, D-50937 K\"{o}ln, Germany}
\affiliation{Horia Hulubei National Institute of Physics and Nuclear Engineering, Bucharest, Romania}

\author{E. Litvinova}
\affiliation{Department of Physics, Western Michigan University, Kalamazoo, MI 49008, USA}
\affiliation{National Superconducting Cyclotron Laboratory, Michigan State University, East Lansing, MI 48824, USA}

\author{C. M{\"u}ller-Gatermann}
\affiliation{Institut f\"{u}r Kernphysik, Universit\"{a}t zu K\"{o}ln, Z\"{u}lpicher Straße 77, D-50937 K\"{o}ln, Germany}

\author{S.G. Pickstone}
\affiliation{Institut f\"{u}r Kernphysik, Universit\"{a}t zu K\"{o}ln, Z\"{u}lpicher Straße 77, D-50937 K\"{o}ln, Germany}

\author{S. Prill}
\affiliation{Institut f\"{u}r Kernphysik, Universit\"{a}t zu K\"{o}ln, Z\"{u}lpicher Straße 77, D-50937 K\"{o}ln, Germany}

\author{P. Scholz}
\affiliation{Institut f\"{u}r Kernphysik, Universit\"{a}t zu K\"{o}ln, Z\"{u}lpicher Straße 77, D-50937 K\"{o}ln, Germany}

\author{A. Zilges}
\affiliation{Institut f\"{u}r Kernphysik, Universit\"{a}t zu K\"{o}ln, Z\"{u}lpicher Straße 77, D-50937 K\"{o}ln, Germany}


\date{\today}

\begin{abstract}

\noindent\textbf{Background:} The semi-magic Sn ($Z = 50$) isotopes have been subject to many nuclear-structure studies. Signatures of shape coexistence have been observed and attributed to two-proton-two-hole (2p-2h) excitations across the $Z = 50$ shell closure. In addition, many low-lying nuclear-structure features have been observed which have effectively constrained theoretical models in the past. One example are so-called quadrupole-octupole coupled states (QOC) caused by the coupling of the collective quadrupole and octupole phonons. 

\noindent\textbf{Purpose:} Proton-scattering experiments followed by the coincident spectroscopy of $\gamma$ rays have been performed at the Institute for Nuclear Physics of the University of Cologne to excite low-spin states in $^{112}$Sn and $^{114}$Sn, to determine their lifetimes and extract reduced transitions strengths $B(\Pi L)$.

\noindent\textbf{Methods:} The combined spectroscopy setup SONIC@HORUS has been used to detect the scattered protons and the emitted $\gamma$ rays of excited states in coincidence. The novel $(p,p'\gamma)$ DSA coincidence technique was employed to measure sub-ps nuclear level lifetimes.

\noindent\textbf{Results:} 74 level lifetimes $\tau$ of states with $J = 0 - 6$ were determined. In addition, branching ratios were deduced which allowed the investigation of the intruder configuration in both nuclei. Here, $sd$ IBM-2 mixing calculations were added which support the coexistence of the two configurations. Furthermore, members of the expected QOC quintuplet are proposed in $^{114}$Sn for the first time. The $1^-$ candidate in $^{114}$Sn fits perfectly into the systematics observed for the other stable Sn isotopes.  

\noindent\textbf{Conclusions:} The $E2$ transition strengths observed for the low-spin members of the so-called intruder band support the existence of shape coexistence in $^{112,114}$Sn. The collectivity in this configuration is comparable to the one observed in the Pd nuclei, i.e. the 0p-4h nuclei. Strong mixing between the $0^+$ states of the normal and intruder configuration might be observed in $^{114}$Sn. The general existence of QOC states in $^{112,114}$Sn is supported by the observation of QOC candidates with $J \neq 1$.

\end{abstract}


\maketitle


\section{Introduction}

The low-energy and low-spin level scheme of the semi-magic stable Sn isotopes has been considered as a ``textbook'' example of the seniority scheme, see, {\it e.g.}, Ref.\,\cite{Tal93a}. However, shell-model calculcations with a finite-range force pointed out that at least configurations with two broken pairs, i.e. seniority $\nu \leq 4$ are needed to fully account for the low-energy levels\,\cite{Bon85a}. To describe the excitation energy of the $3^-_1$ state, one particle - one hole neutron configurations had to be included which originated from excitations across the $^{100}$Sn inert core. In addition to these structures, low-energy two proton - two hole intruder states are observed in the Sn isotopes, see, {\it e.g.}, the review articles\,\cite{Wood92a, Heyd11a} and references therein. These positive-parity states will likely mix with states of the ``normal'' configuration. It is, thus, not trivial whether rather pure collective quadrupole states of two- and three-phonon nature will be observed in the Sn isotopes. Experimental studies in $^{124}$Sn have identified candidates for members of the three-phonon multiplet\,\cite{Band05a}. In fact, identifying such structures in the semi-magic Sn nuclei has been named as an important step to answer the question as to whether pure vibrational modes can be observed in the $Z = 50$ region\,\cite{Garr10a}. 
\\

For years the Cd isotopes had been considered as prime examples exhibiting the vibrational character put forward by Bohr and Mottelson\,\cite{Bohr75a}. But this simple picture has been questioned\,\cite{Garr10a} also due to the existence of shape coexistence at low energies in the Cd isotopes, see, {\it e.g.}, the recent review article\,\cite{Garr16a} and references therein. In addition, a quintuplet of negative-parity states is expected due to the coupling of the collective quadrupole and octupole phonon in vibrational nuclei. Its study might further help to understand the concept of vibrational excitations in nuclei since Pauli blocking is expected to be smaller and since these states will not mix with positive-parity intruder states. The $1^-$ member of this multiplet has been studied systematically in $^{112,116-124}$Sn\,\cite{Bryss99a, Pys06a}.
\\

Many lifetimes in $^{112}$Sn are known from an $(n,n'\gamma)$ experiment performed at the University of Kentucky (USA)\,\cite{Kum05a, Orce07a}. These inelastic neutron-scattering (INS) experiments use the $^{3}$H$(p,n){}^{3}$He reaction to generate neutrons of different energies by tuning the incident proton energy. To minimize feeding effects, several neutron energies are usually used to extract lifetimes with the INS Doppler-shift attenuation (DSA) technique. The scattered neutrons are, however, not detected in coincidence with the $\gamma$ rays. For more details on this method, see, {\it e.g.}, Refs.\,\cite{Pet13a,Kum05a} and references therein. In contrast, the new $(p,p'\gamma)$ DSA coincidence technique with the SONIC@HORUS setup\,\cite{Pick17a} at the University of Cologne (Germany) detects the scattered protons in coincidence with the $\gamma$-rays emitted from the excited state to determine nuclear level lifetimes without feeding contributions\,\cite{Henn15a}. A further advantage of this method is that much less target material is needed compared to the $(n,n'\gamma)$ experiments. Thus, also less abundant isotopes such as $^{114}$Sn can be studied with this method.
\\ 

It is the purpose of this work to report on the two $(p,p'\gamma)$ experiments we performed to study excited states in $^{112,114}$Sn up to an excitation energy of 4\,MeV and determine the level lifetimes.

\section{experimental details}

The $^{112,114}$Sn$(p,p'\gamma)$ experiments were performed at the 10\,MV FN-Tandem facility of the University of Cologne where the protons were accelerated to an energy of $E_p = 8$\,MeV. The combined spectroscopy setup SONIC@HOR\-US used for the $p \gamma$-coincidence experiments consists of passivated implanted planar silicon (PIPS) detectors and up to 14 high-purity germanium (HPGe) detectors. Six of these can be equipped with BGO shields for active Compton suppression\,\cite{Pick17a}. The current on target accounted to about 5\,nA with a master-trigger rate of up to 25\,kHz, which corresponded to operating the silicon and germanium detectors at maximum count rates of about 11\,kHz and 15\,kHz, respectively. $p \gamma$-coincidence data for excited states up to about 4\,MeV were acquired by using XIA's DGF-4C Rev.\,F modules\,\cite{Hubb99a, Skul00a, Warb06a}. A level-2 global first level trigger (GFLT) was set externally to record twofold and multifold coincidences as listmode data\,\cite{Elv11a, Henn14a}. The SONIC chamber, i.e. SilicON Identification Chamber, which was used in these experiments was the second SONIC version housing seven silicon detectors in total\,\cite{Pick17a}. Four of these detectors are placed in tubes at angles of $\theta = \SI{122}{\degree},~ \phi = \SI{45}{\degree}, \SI{135}{\degree}, \SI{225}{\degree}, \SI{315}{\degree}$, while another three silicon detectors can be fixed to the chamber by using magnets at $\theta = \SI{114}{\degree},~ \phi = \SI{0}{\degree}, \SI{90}{\degree}$ and $\SI{180}{\degree}$, respectively.
\\

\begin{figure}[t]
\centering
\includegraphics[width=0.75\linewidth]{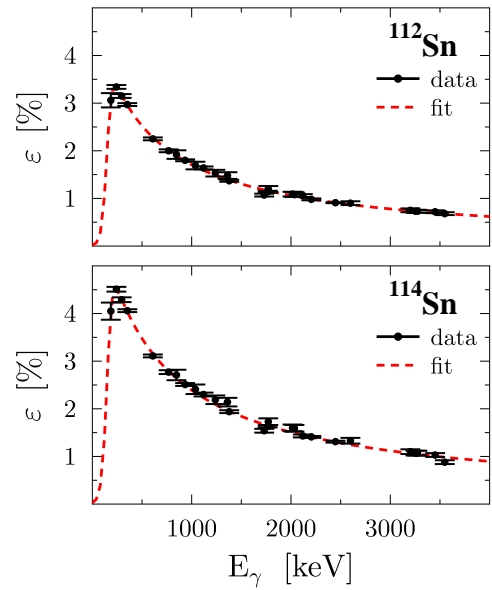}
\caption{\label{fig:efficiency}(color online) Absolute photopeak efficiencies for the $^{112,114}$Sn$(p,p'\gamma)$ experiments. The empirical Wiedenh{\"o}ver function (red dashed line) has been used to fit the experimentally measured efficiencies (black circles)\,\cite{Wieden94a}.}
\end{figure}

A precise energy calibration of the HPGe detectors is crucial for any DSA lifetime measurement. For the $(p,p'\gamma)$ DSA coincidence technique a $^{56}$Co calibration source is mounted in SONIC throughout the experiment to guarantee for this precise calibration. The energy calibration of the HPGe detectors is precise to a level of at least 0.2\,keV and, thus, $\gamma$-energy centroid shifts which are well below 1\,keV can be recognized. The energy calibration of the silicon detectors is performed by identifying specific excited states of the target nucleus in the proton spectra. The assignments in the proton spectra are cross checked by setting a gate onto $\gamma$ transitions in the HPGe detectors. The absolute photopeak efficiency of the setup was determined using a $^{226}$Ra source of known activity and a $^{56}$Co source whose activity was used as a scaling factor to obtain agreement with the $^{226}$Ra data, see Fig.\,\ref{fig:efficiency}. It is, thus, known up to 3.5\,MeV and no significant extrapolation was needed to study excited states in $^{112,114}$Sn up to an energy of about 4\,MeV. 

\section{Lifetime determination}

\subsection{$\gamma$-energy centroid shifts and the Doppler-shift attenuation factor $F \left( \tau \right)$}

Correlated $p \gamma$-coincidence matrices can be generated from the $p \gamma$-coincidence listmode data. For the case of the DSAM technique, these coincidence matrices can be determined unambiguously since the kinematics of the $(p,p'\gamma)$ allow clear correlations, see, {\it e.g.}, Refs.\,\cite{Henn14a, Henn14b, Henn15a, Henn15b}. In total, there are three of such kinematically correlated DSA groups for excited states  up to 4\,MeV in $^{112,114}$Sn when using the second SONIC version. Each group typically contains eleven subgroups characterized by their common Doppler angle $\Theta$, respectively. If statistics are not sufficient less subgroups can be considered, i.e. more Doppler angles $\Theta$ can be grouped into one subgroup resulting in larger overall $\cos(\Theta)$ uncertainties.
\\

\begin{figure}[!t]
\centering
\includegraphics[width=1\linewidth]{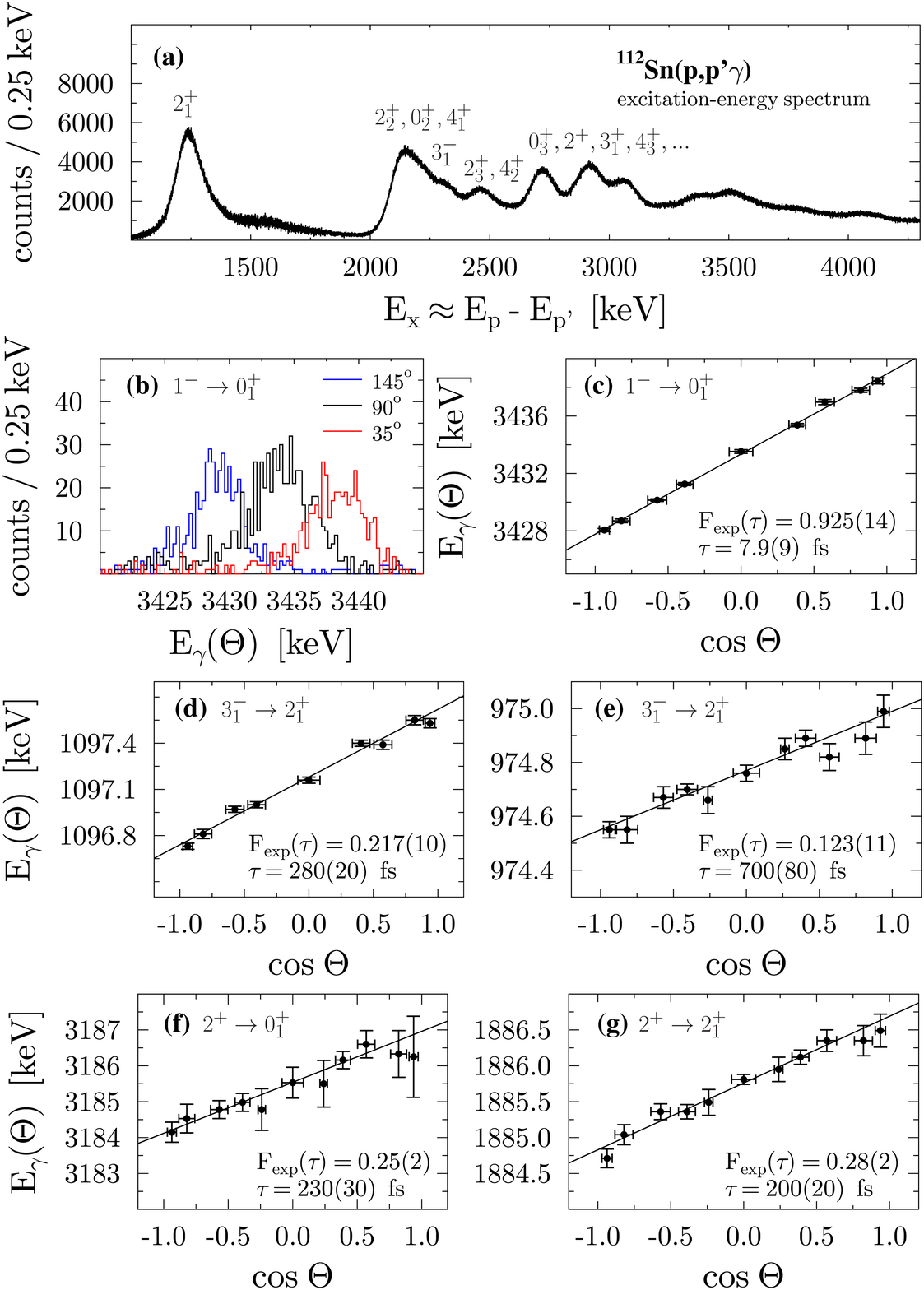}
\caption{\label{fig:Sn_shifts}(color online) Method to extract $\gamma$-energy centroid shifts. {\bf (a)} Excitation-energy spectrum, which corresponds roughly to the energy loss of the incident protons. Several excited states in $^{112}$Sn are marked. Gates can be applied to select the excitation of specific states. {\bf (b)} The $\gamma$-energy centroid shifts are observed in the excitation-energy gated $\gamma$ spectra, {\it e.g.}, the DSAM subgroups at $\Theta = \SI{35}{\degree}, \SI{90}{\degree},$ and $\SI{145}{\degree}$, {\bf (c)} from which a linear dependence on $\cos(\Theta)$ is extracted. The example of the $J^{\pi} = 1^-$ state at $E_x = 3433.4(2)$\,keV in $^{112}$Sn is presented. {\bf (d)}, {\bf (e)} observed energy-centroid shifts of the $3^-_1$ states in $^{112,114}$Sn which are below 1\,keV. {\bf (f)}, {\bf (g)} $\gamma$-energy centroid shifts for $\gamma$-decays to different final states of the $2^+$ at 3185.5(2)\,keV in $^{114}$Sn. The Doppler-shift attenuation factors $F(\tau)$ and the lifetimes $\tau$ determined from a comparison to a Monte-Carlo simulation are also shown in panels {\bf (c)} to {\bf (g)}.}
\end{figure}

Excitation-energy gates select specific excitation regions and exclude feeding contributions. Figs.\,\ref{fig:Sn_shifts}\,{\bf (b)}-{\bf (e)} present the observed energy-centroid shifts of the $1^-$ state at $E_x = 3433$\,keV in $^{112}$Sn and of the $3^-_1$ states in $^{112,114}$Sn, respectively. From the slope of the linear fit, the Doppler-shift attenuation factor $F(\tau)$ can be determined: 

\begin{align}
E_{\gamma}(\Theta,t) = E_{\gamma}^0 \left( 1+ F(\tau) \frac{v_0}{c} \cos \Theta \right) \nonumber
\label{eq:DSAM_03}
\end{align}

If $\gamma$-decay branching is observed, the Doppler-shift attenuation factor can be determined from $\gamma$-decays to different final states. Figs.\,\ref{fig:Sn_shifts}\,{\bf (f)} and {\bf (g)} present the case of a $2^+$ state at 3185\,keV in $^{114}$Sn. As can been seen from the figure, the two $\gamma$-decay branches yield consistent Doppler-shift attenuation factors $F(\tau)$ within their statistical uncertainties.

\subsection{Simulation of the stopping process}

\begin{figure}[t]
\centering
\includegraphics[width=1\linewidth]{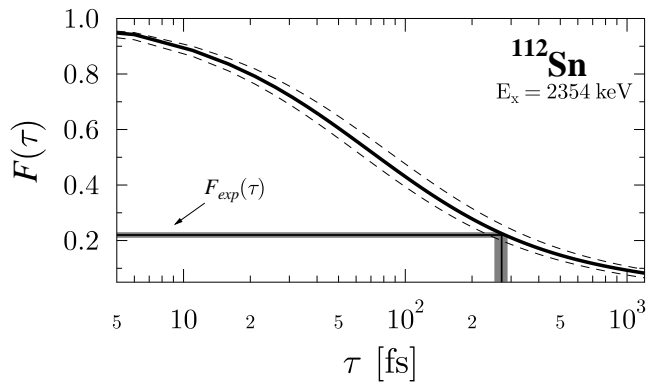}
\caption{\label{fig:ftau}Determination of the lifetime $\tau_{3^-_1}$ from the $F(\tau)$ curve in $^{112}$Sn. The grey band corresponds to the statistical uncertainties of the experimentally determined $F_{\mathrm{exp}}(\tau)$ value. The final lifetime $\tau$ of 280(20) fs is obtained by comparison to the Monte-Carlo simulation. The solid black line corresponds to the calculation using the stopping-power parameters of Table\,\ref{tab:targets}, whereas the dashed lines correspond to the variation of these parameters to estimate the systematic uncertainties.}
\end{figure}

Nuclear level lifetimes $\tau$ were determined from a comparison of the experimentally extracted $F(\tau)$ values with the predictions of a Monte-Carlo simulation of the stopping process, see Fig.\,\ref{fig:ftau} and Ref.\,\cite{Henn15a}. The Monte-Carlo simulation considers electronic and nuclear stopping according to the formalism of Lindhard, Scharff, and Schi\o{}tt (LSS)\,\cite{Lind63a, Lind68a}. The computer code\,\cite{Pet15a}, which has been used in this work, is a modified version of the DESASTOP program by Winter\,\cite{Wint83a, Wint83b} where the universal scattering function described in Ref.\,\cite{Curr69a} has been implemented. It has been further modified to be able to deal with multi-layered target compositions, i.e. especially alloy layers of different composition and also allows the implementation of details of the experimental setup to obtain a more constrained simulation. Besides the stopping powers in the target and stopper material, the areal densities of the materials have to be known as precisely as possible. Otherwise, severe errors in the lifetime calculation might be introduced as shown in, {\it e.g.}, Refs.\,\cite{Pet13a}.
\\

\begin{figure}[t]
\centering
\includegraphics[width=0.75\linewidth]{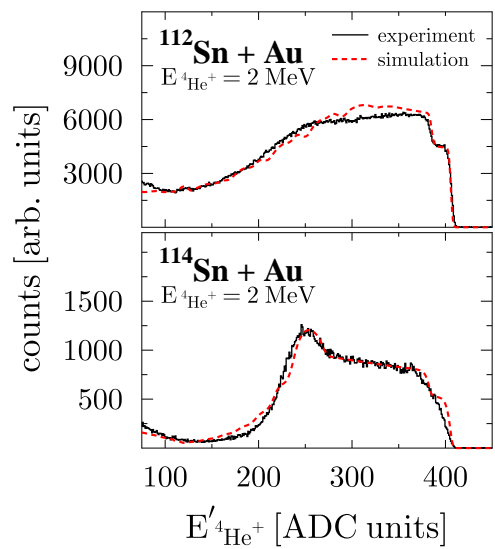}
\caption{\label{fig:Sn_rbs}(color online) RBS spectra for the $^{112,114}$Sn\,+\,Au foils (black line) and the simulation (red dashed line) using the SIMNRA software\,\cite{SIMNRA1, SIMNRA2}. See the text for more details.}
\end{figure}

Unfortunately, Rutherford Backscattering Spectrometry (RBS) experiments with 2\,MeV $^{4}$He$^+$ ions, performed at the RUBION facility of the Ruhr-Universit{\"a}t Bochum in Germany, revealed that a Sn-Au alloy had formed, i.e. the target and stopper material were not completely separated. However, by using the RBS simulation software SIMNRA\,\cite{SIMNRA1, SIMNRA2} and by introducing a thickness distribution in the different alloy layers, a reasonable description of the experimental RBS spectra could be obtained, see Fig.\,\ref{fig:Sn_rbs}. Thus, the areal densities and the relative contributions of Sn and Au to the different layers could be extracted.

\begin{table}[t] 
\centering
\caption{\label{tab:targets}Target properties and stopping-power parameters $f_e$ and $p$ for the $^{112,114}$Sn+Au foils. $f_n$ has been set to 0.85, see text. The different layer densities were determined from the RBS analysis. The respective stopping powers were determined by taking into account the layer compositions.}
\begin{ruledtabular}
\begin{tabular}{ccccccc}
Layer & $a_{\mathrm{Sn}}$ & $b_{\mathrm{Au}}$ & density alloy & areal density & $f_e$ & $p$ \\
      &    &    & [g/cm$^3$]    & [mg/cm$^2$]   &       &     \\
\hline
\multicolumn{7}{c}{{\bf $^{112}$Sn+Au alloy}} \\
\hline
1 & 0.50 & 0.50 & 13.1 & 0.87 & 0.61 & 0.61 \\
2 & 0.64 & 0.36 & 11.4 & 0.14 & 0.64 & 0.60 \\
\multicolumn{7}{c}{$\rightarrow$ total areal density\,=\,1.01\,mg/cm$^2$} \\
\hline
\multicolumn{7}{c}{{\bf $^{114}$Sn+Au alloy}} \\
\hline
1 & 0.55 & 0.45 & 12.7 & 0.70 & 0.63 & 0.61 \\
2 & 0.37 & 0.63 & 14.9 & 0.13 & 0.59 & 0.61 \\
3 & 0.42 & 0.58 & 14.3 & 0.14 & 0.60 & 0.61 \\
\multicolumn{7}{c}{$\rightarrow$ total areal density\,=\,0.97\,mg/cm$^2$} \\
\end{tabular}
\end{ruledtabular}
\end{table}

The alloy introduces some additional considerations which have to be made. First of all, the stopping powers for tin recoils in the alloy results as the sum of the relative contributions to the respective layer $i$, i.e.

\begin{align}
\left(\frac{dE}{dx}\right)_{\mathrm{alloy},i} = \left[ a \cdot \left(\frac{dE}{dx}\right)_{\mathrm{Sn}} + b \cdot \left(\frac{dE}{dx}\right)_{\mathrm{Au}} \right]_{i}, \nonumber
\end{align} 

which are determined from the stopping-power tables of Northcliffe and Schilling\,\cite{North70a} and corrections to these due to electronic structures of the respective material\,\cite{Zieg74a}. Here, $a$ and $b$ are known from the RBS data simulation. Second, effective charges $Z$ and effective masses $A$ are introduced to transform $E$ and $x$ to the dimensionless variables of the LSS theory\,\cite{Curr69a} by also using $a$ and $b$. And third, one introduces compound densities, which are not necessarily homogeneous and are not compatible with a simple averaging. The stopping power results including the RBS analysis are shown in Table\,\ref{tab:targets}. Furthermore, the $^{112}$Sn and $^{114}$Sn targets were only enriched to 85.5\,$\%$ and 66.5\,$\%$, respectively. This had to be taken into account in the Monte-Carlo simulation since the density profile is affected. Due to these complications, a thorough check of the input parameters for the Monte-Carlo simulations by means of known lifetimes was necessary. Results will be presented in Sec.\,\ref{sec:tau_comp}. 

\begin{figure}[!t]
\centering
\includegraphics[width=0.9\linewidth]{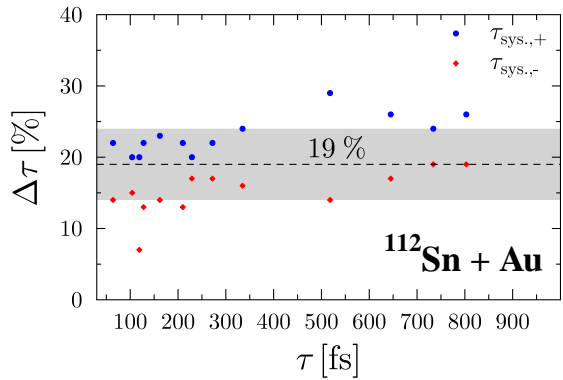}
\caption{\label{fig:sysun}(color online) Systematic uncertainties of the lifetime measurements. The lifetime $\tau$ has been calculated with the stopping power parameters of Table\,\ref{tab:targets}. The systematic uncertainties correspond to the variation of the lifetime $\tau$ in percent when these parameters are changed as described in the text. A mean uncertainty of about 19\,$\%$ is calculated.}
\end{figure}

Since the lifetime measurement relies on the correct determination of the stopping powers, systematic uncertainties should be estimated as well. To do so, the $f_e$ parameter of the electronic stopping power was varied by 10\,$\%$ since the fit to the tabulated electronic stopping powers of Northcliffe and Schilling\,\cite{North70a} is characterized by a 10\,$\%$ uncertainty. Furthermore, calculations were performed with different screening factors $f_n$ for the nuclear potential of $f_n = 0.7, 0.85$, and $1.0$, i.e. approximately a 18\,$\%$ variation. The $f_n = 1.0$ value is the standard value of the LSS theory\,\cite{Curr69a}, 0.85 has been used in this work, and $f_n = 0.7$ is commonly used. Within this parameter range, the combinations of $f_e$ and $f_n$ variation in the different layers leading to the largest variation of the lifetime $\Delta \tau_{sys.} = |\tau_{\mathrm{sys.,}\pm} - \tau|/\tau$ were calculated, where ``+'' indicates a longer lifetime and ``-'' a shorter lifetime. The results are shown in Fig.\,\ref{fig:sysun} for different lifetimes in $^{112}$Sn determined in this work. The systematic uncertainties are conservatively estimated with $\Delta \tau_{sys.,-} \leq 19\,\%$. 
\\

The lifetimes extracted in this work might appear slightly longer compared to previous results, even though there are several exceptions, see Table\,\ref{tab:112Sn_tau}. Due to the low recoil energies in the $(p,p'\gamma)$ reaction, i.e. $E_{\mathrm{rec.}} < 200$\,keV, in contrast to heavy-ion induced reactions, where the line-shape analysis is used, the proton-induced reaction is much more sensitive to the nuclear stopping power, i.e. the screening factor $f_n$\,\cite{Curr69a}. Despite the good agreement, it cannot be excluded that the different alloy layers are assumed too dense in the RBS simulation. This would also result in slightly longer lifetime values and explain the larger $f_n$ value, i.e. the larger nuclear-stopping contribution needed in our analysis to obtain agreement with known lifetimes. At present, no decision in favor of any of these two scenarios can be made due to missing sensitivity to the very details of the target+stopper alloy composition.

\section{experimental results}

In total 74 lifetimes and lifetime limits have been determined. Out of these, 39 lifetimes have been determined for the first time; 13 lifetimes in $^{112}$Sn and 26 lifetimes in $^{114}$Sn, respectively. The lifetimes are given in Tables~\ref{tab:112Sn_tau} and \ref{tab:114Sn_tau}. Note that only statistical uncertainties are stated here. As has been mentioned in the previous section, systematic uncertainties should be considered at the 19\,$\%$ level or smaller. Besides the determination of nuclear level lifetimes, several new and also weak $\gamma$ decays of excited states were observed with SONIC@HORUS which are marked with $*$. Some of these have relative $\gamma$-decay intensities $I_{\gamma}$ of smaller than $1\,\%$. For most cases, the lifetimes $\tau$ and $\gamma$-decay intensities $I_{\gamma}$ are in very good agreement with previously known and adopted values\,\cite{ENSDF}. However, discrepancies are observed. We have observed twelve new excited states and 116 new $\gamma$ transitions in $^{112}$Sn. In $^{114}$Sn, six new levels and 33 new $\gamma$ transitions were found.

\renewcommand*{\arraystretch}{1.2}
\begin{longtable*}{p{0.10\textwidth}p{0.09\textwidth}p{0.09\textwidth}p{0.09\textwidth}p{0.08\textwidth}p{0.09\textwidth}p{.10\linewidth}p{0.09\textwidth}p{0.09\textwidth}p{0.10\textwidth}}
\caption{Experimental data for excited states in $^{112}$Sn. The $\gamma$-decay intensities $I_{\gamma}$, multipole-mixing ratios $\delta$, Doppler-shift attenuation factors $F(\tau)$, lifetimes $\tau$ determined in this work, and the lifetimes $\tau_{\mathrm{lit.}}$ known from literature are given, respectively. The multipole-mixing ratios $\delta$ and lifetimes $\tau$ have been determined in Ref.\,\cite{Kum05a} if not indicated otherwise. Shown are all $\gamma$ decays with their $\gamma$-decay energies and intensities, which have been observed for a given excited state. Newly observed $\gamma$ decays are marked with $*$. Newly observed excited states are marked with $\dagger$. Only statistical uncertainties are given for the lifetime values. As explained in the text, systematic uncertainties should be considered at the 19\,$\%$ level.}
\label{tab:112Sn_tau}

\vspace*{1mm}
\\
\toprule
$E_x$\,[keV]& $J^{\pi}_i$ & $J^{\pi}_f$ & $E_{\gamma}$\,[keV] & $I_{\gamma}$\,[$\%$] & $\delta$ & $\Pi L$ & $F(\tau)$ & $\tau$\,[fs] & $\tau_{\mathrm{lit.}}$\,[fs]\\
\hline

\endfirsthead

\multicolumn{7}{c}{{\small Table \ref{tab:112Sn_tau}: Continuation}} \\

\hline
\hline

$E_x$\,[keV]& $J^{\pi}_i$ & $J^{\pi}_f$ & $E_{\gamma}$\,[keV] & $I_{\gamma}$\,[$\%$] & $\delta$ & $\Pi L$ & $F(\tau)$ & $\tau$\,[fs] & $\tau_{\mathrm{lit.}}$\,[fs]\\
\hline
\endhead
\endfoot
\hline
\hline
\multicolumn{10}{l}{$^{\mathrm{a}}$ Taken from Ref.\,\cite{ENSDF}.}\\
\multicolumn{10}{l}{$^{\mathrm{b}}$ 650(40) fs in Ref.\,\cite{Jungcl11a}.}\\
\multicolumn{10}{l}{$^{\mathrm{c}}$ Taken from Ref.\,\cite{Pys06a}.}\\
\multicolumn{10}{l}{$^{\mathrm{d}}$ Taken from Ref.\,\cite{Oez14a}.}
\endlastfoot

1256.5(2)& $2^+_1$ & $0^+_1$ & 1256.5(2)& 100 & -& $E2$ & 0.110(6) & 800(110) & 542(7)$^{\mathrm{a,b}}$\\
\hline	
2150.5(3)& $2^+_2$ & $0^+_1$ & 2150.5(2)& 20(3) & -& $E2$ & \multirow{3}{*}{$< 0.04$}  & \multirow{3}{*}{$> 2700$} & \multirow{3}{*}{2000(580)$^{\mathrm{a}}$} \\
&\multirow{2}{*}{$2^+_2$} & \multirow{2}{*}{$2^+_1$} & \multirow{2}{*}{893.9(2)} & \multirow{2}{*}{100} & -0.28(6)& \multirow{2}{*}{$M1+E2$} &  &  &\\
&  &  & &  & $7^{+3}_{-2}$& &  &  & \\
\hline	
2190.5(2)& $0^+_2$ & $2^+_1$ & 934.0(2) & 100 & -& $E2$ &  &  & $\geq 3900$\\
\hline
2247.0(3) & $4^+_1$ & $2^+_1$ & 990.47(10) & 100 &  & $E2$ & $< 0.03$ & $> 4400$& 4800(720)$^{\mathrm{a}}$\\
\hline	
2353.7(2)& $3^-_1$ & $2^+_1$ & 1097.2(2)& 100 & 0.02(2) & $\centering E1$ &  0.217(10) & 280(20) & 310(20)$^{\mathrm{a}}$\\	
\hline
2475.5(2)& $2^+_3$ & $0^+_1$ & 2475.5(2)& 100 & - & $E2$ & 0.049(6) & \multirow{2}{*}{2500(400)} & \multirow{2}{*}{$>$\,3500} \\
& $2^+_3$ & $2^+_1$ & 1218.9(2)& 36(5) & -0.54(7) & $M1+E2$ & 0.039(12)  &  & \\
& $2^+_3$ & $0^+_2$ & 284.9(2)& 0.70(10)& - & $E2$&  &  &\\
\hline
2520.5(2)& $4^+_2$ & $2^+_1$ & 1264.0(2)& 100 & -0.04(4) & $E2$ & 0.067(9) & 1600(300) & $>$\,1100\\
\hline
2548.6(2)& $6^+_1$ & $4^+_1$ & 301.6(2)& 100 & & $(E2)$ & & &19.81(12)\,ns$^{\mathrm{a}}$\\
\hline
2617.4(3)& $0^+_3$ & $2^+_1$ & 1360.9(3)& 100 & - & $E2$ & $< 0.03$& $> 4200$ & $> 580$\\
         & $0^+_3$ & $2^+_2$ & 466.8(4)$^*$& 1.2(3) & - & $E2$ &  & & \\
\hline
2720.6(2) & $2^+_4$ & $0^+_1$ & 2720.6(2)& 10.0(14) & - & $E2$& & &\\
          & $2^+_4$ & $2^+_1$ & 1464.1(2)& 100 & 0.17(10) & $M1+E2$ & 0.066(9) & 1500(300) & 1100$^{+\,1500}_{-\,450}$\\
          & $2^+_4$ & $2^+_2$ & 570.0(2)$^*$& 1.4(2) & & & &\\
          & $2^+_4$ & $0^+_2$ & 529.7(3)$^*$& 1.0(2) & - & $E2$ & & &\\
          & $2^+_4$ & $3^-_1$ & 366.6(3)$^*$& 13(2)& & $(E1)$ & &\\
\hline
2755.2(3) & $3^+_1$ & $2^+_1$ & 1499.0(3) & 25(4) & 0.03(5) & $M1 + E2$ & & & $> 1150$\\ 	
          & $3^+_1$ & $2^+_2$ & 604.8(2) & 4.4(7) & & & & &\\
          & $3^+_1$ & $4^+_1$ & 508.3(2) & 100 & 0.2(1) & & & &\\
          & $3^+_1$ & $3^-_1$ & 401.5(3) & 0.8(2) & & & & &\\
          & $3^+_1$ & $4^+_2$ & 234.7(6) & 1.4(3) & & & & &\\
\hline
2764.9(2)& $< 5$ & $2^+_1$ & 1508.5(2)& 100 & & &  0.055(8) & 2400(700) & $>$\,1500\\
         & $< 5$ & $2^+_2$ & 614.3(3)$^*$& 1.9(5) & & &   &  &\\	
\hline
2783.5(2)& $4^+$ & $2^+_1$ & 1527.0(2)& 100 & -0.06(4) & $E2$ &  0.127(10) & 580(70) & 450$^{+\,150}_{-\,90}$ \\
         & $4^+$ & $4^+_1$ & 536.3(3)$^*$& 2.7(5) & & &  &  & \\
\hline
2913.0(5)& $4^+$ & $2^+_1$ & 1656.2(2)& 91(13) & -0.11(11) & $E2$ &  &  &\\
         & $4^+$ & $2^+_2$ & 762.5(2)$^*$ & 7(2) &  & $(E2)$ &  &  & \\
         & $4^+$ & $4^+_1$ & 665.4(2) & 100 &  &  &  0.11(4) & \multirow{2}{*}{420(140)}  & \multirow{2}{*}{$> 940$}\\
         & $4^+$ & $3^-_1$ & 559.1(2)$^*$ & 14(2) &  & $(E1)$ & 0.18(5)  &  &\\
\hline
2917.0(2)& $(2^+,3,4^+)$ & $2^+_1$ & 1660.5(3)$^*$& 2.1(4) &  &  &  &  & $> 1600$\\
         & $(2^+,3,4^+)$ & $2^+_2$ & 766.4(2) & 8.5(12) &  &  &  &  &\\
         & $(2^+,3,4^+)$ & $4^+_1$ & 669.8(2) & 100 &  &  &  &  &\\
         & $(2^+,3,4^+)$ & $4^+_2$ & 396.8(3)$^*$ & 0.8(2) &  &  &  &  &\\
\hline
2926.6(4)& $6^+_2$ & $6^+_1$ & 377.4(2) & 100 &  &  & 0.50(12)  & 80(40) & $> 300$\\
\hline
2945.0(7)& $4^+$ & $2^+_1$ & 1688.5(2)& 100 &  & $(E2)$ & 0.036(10) & 3100(1000) & $> 1600$\\
         & $4^+$ & $2^+_2$ & 794.2(2) & 5.4(10) &  & $(E2)$ &  &  &\\
         & $4^+$ & $4^+_1$ & 697.9(2)$^*$ & $< 1.5$ &  &  &  &  &\\
         & $4^+$ & $2^+_3$ & 469.5(2) & 18(3) &  & $(E2)$ &  &  &\\
         & $4^+$ & $4^+_2$ & 424.6(3)$^*$ & 4.9(9) &  & &  &  & \\
         & $4^+$ & $6^+_1$ & 396.4(4)$^*$ & 2.3(5) &  & $(E2)$  &  &  & \\
         & $4^+$ & $4^+$ & 161.4(2)$^*$ & 9(2) &  &  &  &  &\\
\hline
2966.4(3)& $2^+$ & $0^+_1$ & 2966.4(4)& 61(9) & - & $E2$ & 0.061(11) & \multirow{2}{*}{1600(300)}  & \multirow{2}{*}{$660^{+1200}_{-280}$}\\
		 &  $2^+$& $2^+_1$ & 1709.7(3)& 100 & 0.3(4) & $M1 + E2$ & 0.069(13) & &\\	 
		 &  $2^+$& $4^+_1$ & 718.0(3)$^*$& 3.8(6)& & $(E2)$ & & &\\	
		 &  $2^+$& $3^-_1$ & 612.3(2)& 29(4)& & $(E1)$ & & &\\	
\hline
2969.0(2)& $(1,3)$ & $2^+_1$ & 1712.5(2)& 100 & & &  0.131(12) & 610(70) & 430$^{+\,300}_{-\,130}$\\
		 &  $(1,3)$& $2^+_2$ & 818.3(2) & 7.5(11)& & & & &\\
\hline
2985.7(2) & $0^+$ & $2^+_1$ & 1729.2(2) & 100 & - & $E2$&  0.086(11) & 1100(190) & $>$\,2400\\	
		 &  $0^+$& $2^+_2$ & 835.3(2)$^*$ & 1.0(3)& - & $E2$ & & &\\
\hline
3077.8(3) & $3^+$ & $2^+_1$ & 1821.6(3) & 83(12) & -1.3$^{+0.3}_{-0.5}$ & $M1+E2$ & & & $> 1800$\\
          & \multirow{2}{*}{$3^+$} & \multirow{2}{*}{$2^+_2$} & \multirow{2}{*}{927.5(2)} & \multirow{2}{*}{100} & 3.0(10) & \multirow{2}{*}{$M1+E2$} & & &\\
          & &  &  & & 0.60$^{+0.10}_{-0.20}$ & & & &\\
          & $3^+$ & $4^+_1$ & 831.1(4) & 9(2) & & & & &\\
          & $3^+$ & $2^+_3$ & 601.8(5)$^*$ & 9(2) & & & & &\\
          & $3^+$ & $4^+_2$ & 557.3(3) & 3.8(10) & & & & &\\
          & $3^+$ & $2^+$ & 357.2(2)$^*$ & 10(2) & & & & &\\
          & $3^+$ & $3^+_1$ & 322.5(2)$^*$ & 10(2) & & & & &\\
\hline
3092.4(2) & $2^+$ & $0^+_1$ & 3092.4(2) & 29(4) & - & $E2$ &  0.12(2) & \multirow{2}{*}{530(50)} & \multirow{2}{*}{360$^{+\,110}_{-\,70}$}\\
		  & $2^+$ & $2^+_1$ & 1835.8(2) & 100 & -1.5(10) & $M1+E2$ &  0.138(11) & & \\
		  & $2^+$ & $0^+_2$ & 901.9(3)$^*$ & 1.6(3) & - & $E2$ & & &\\ 	
\hline
3113.2(2) & $(2^+,3,4^+)$ & $2^+_1$ & 1856.8(2)$^*$ & 4.0(7) & & & & &\\
          & $(2^+,3,4^+)$ & $2^+_2$ & 962.7(2) & 100 & & & 0.06(3) & 1600(1000)& -\\  
          & $(2^+,3,4^+)$ & $4^+_1$ & 866.0(2)$^*$ & 17(2) & & & & & \\  
          & $(2^+,3,4^+)$ & $3^-_1$ & 759.2(2)$^*$ & 3.5(6) & & & & & \\
          & $(2^+,3,4^+)$ & $2^+_3$ & 637.7(2)$^*$ & 2.1(5) & & & & & \\
          & $(2^+,3,4^+)$ & $2^+_4$ & 392.6(2)$^*$ & 16(2) & & & & & \\
          & $(2^+,3,4^+)$ & $3^+_1$ & 357.2(3)$^*$ & 4.7(8) & & & & & \\        
          & $(2^+,3,4^+)$ & $4^+_3$ & 329.6(2)$^*$ & 15(2) & & & & &\\  
          & $(2^+,3,4^+)$ & $4^+_4$ & 200.5(4)$^*$ & 1.2(3) & & & & &\\  
\hline
3132.5(2) & $5^-_1$ & $4^+_1$ & 885.6(2) & 100 & -0.02$^{+0.01}_{-0.04}$& $E1$ & & & $> 1450$\\
          & $5^-_1$ & $3^-_1$ & 778.7(4) & 18(7) & & $(E2)$ & & &\\  
\hline
3148.3(2) & $4^+$& $2^+_1$ & 1891.8(2) & 100 & 0.05(10) & $E2$ & & & 800$^{+\,1400}_{-\,300}$\\
\hline
3248.2(2) & $2^+$ & $0^+_1$ & 3248.2(2) & 100 & - & $E2$ &  0.072(10) & 1400(300) & $>$\,1600\\ 	
          & $2^+$ & $2^+_1$ & 1991.7(2) & 21(3)& & & & & \\ 
          & $2^+$ & $3^-_1$ & 894.0(2) & $< 82$ & & $(E1)$ & & &\\ 
          & $2^+$ & $2^+_3$ & 772.7(2) & 15(2)& & & & &\\ 
\hline
3272.5(2) & $4^+$ & $2^+_1$ & 2016.0(2) & 100 & -0.0(1)& $E2$ &  0.11(2) & 730(200) & 430$^{+\,300}_{-\,130}$\\
          & $4^+$ & $2^+_2$ & 1122.1(2) & 2.5(8) & & $E2$ & & &\\
          & $4^+$ & $4^+_1$ & 1025.6(2)$^*$ & 13(2) & & & & &\\
          & $4^+$ & $6^+_1$ & 723.7(3)$^*$ & 1.9(7) & & $(E2)$ & & &\\
          & $4^+$ & $4^+$ & 488.9(4)$^*$ & 2.8(11) & & & & &\\
\hline
3285.7(2) & $2^+$ & $0^+_1$ & 3285.7(2) & 100 & - & $E2$ &  0.19(2) & \multirow{2}{*}{290(30)}  & \multirow{2}{*}{320$^{+\,220}_{-\,100}$}\\
          & $2^+$ & $2^+_1$ & 2029.2(2) & 82(14) & & &  0.22(2) & &\\ 	
          & $2^+$ & $2^+_2$ & 1135.5(3)$^*$ & 4.3(8) & & & & &\\ 
          & $2^+$ & $3^-_1$ & 931.9(3)$^*$ & 6(2) & & $(E1)$ & & &\\ 
\hline
3337.8(2) & $2^+$ & $2^+_1$ & 2081.3(2) & 100 & & &  0.15(2) & 470(90) & $>$\,480 \\
          & $2^+$ & $2^+_2$ & 1187.3(2)$^*$ & 20(3) & & & & \\ 	
\hline
3352.8(2) & $2^+$ & $0^+_1$ & 3352.8(2) & 100 & - & $E2$ &  0.055(12) & 2600(700) & $>$\,2000 \\
          & $2^+$ & $2^+_1$ & 2096.3(2) & 30(4) & & & & \\ 	
          & $2^+$ & $2^+_2$ & 1202.5(2)$^*$ & 23(3) & & & &\\
          & $2^+$ & $4^+_1$ & 1105.7(3)$^*$ & 4.7(8) & & $(E2)$ & & \\
          & $2^+$ & $0^+_3$ & 735.4(5)$^*$ & 1.7(6) & - & $E2$ & &\\
          & $2^+$ & $2^+$ & 631.7(3)$^*$ & 6.0(11) & & & &\\
\hline
3378.3(2) & (1,$2^+$) & $2^+_1$ & 1227.8(2) & 100 & & & & & \\
          & (1,$2^+$) & $2^+_3$ & 903.0(3)$^*$ & 6(2) & & & & & \\
\hline
3383.3(2) & $3^-$ & $2^+_1$ & 2126.8(2) & 100 & 0.1(5) & $E1$ &  0.197(14) & 310(20) & 260$^{+\,120}_{-\,70}$\\
          & $3^-$ & $2^+_2$ & 1232.9(2)$^*$ & 4.8(11) & & & &\\
          & $3^-$ & $(2^+,3,4^+)$ & 466.5(2) & 13(2) & & & &\\
\hline
3396.6(2) & $2^{(-)}$ & $2^+_1$ & 2139.9(2)$^*$ & 9(2) & & & & & \\
          & $2^{(-)}$ & $2^+_2$ & 1246.1(2) & 100 & & &  0.15(3) & 460(130) & 330$^{+\,140}_{-\,80}$ \\
          & $2^{(-)}$ & $3^-_1$ & 1042.4(2) & 42(8) & 1.8(12) & & & &\\
          & $2^{(-)}$ & $2^+$ & 675.8(2)$^*$ & 6.1(13) & & & & &\\
\hline
3412.7(2) & $6^+$ & $4^+_1$ & 1165.7(2) & 100 &  & $(E2)$ &  & & \\
\hline
3417.1(2) & $4^+$ & $2^+_1$ & 2160.6(2) & 100 &  & $(E2)$ &  $< 0.05$ & $> 2216$ & $> 500$ \\
          & $4^+$ & $4^+_1$ & 1170.1(2)$^*$ & 15(3) &  &  &  & &\\
\hline
3433.4(2) & $(1^-)$ & $0^+_1$ & 3433.4(2) & 100 & - & $E1$ &  0.925(14) & 7.9(9) & 4.3(5)$^{\mathrm{c}}$ \\ 	
\hline
3455.7(2) & $2^+,3^+$ & $2^+_1$ & 2199.1(2) & 100 & 2.8(10) & $M1+E2$ & & & $> 940$\\
          & $2^+,3^+$ & $2^+_2$ & 1305.3(2)$^*$ & 27(4) & & & & & \\
          & $2^+,3^+$ & $3^+_1$ & 700.2(2) & 33(5) & & & & & \\
          & $2^+,3^+$ & $2^+$ & 489.5(2)$^*$ & 21(3) & & & & & \\
\hline
3497.9(2) & $5^-$ & $3^-_1$ & 1144.2(2) & 100 & & &  0.17(4) & 410(140) & 64$^{+\,64}_{-\,30}$ \\ 	
          & $5^-$ & $4^+$ & 977.1(2) & 38(8) & & & & &\\
          & $5^-$ & $4^+$ & 714.7(3)$^*$ & $< 5$ & & & & &\\
\hline
3518.6(4)     & $2^+$ & $2^+_2$ & 1368.4(2) & 100 & & &  0.20(4) & 310(90) & -  \\
              & $2^+$ & $0^+_2$ & 1328.2(3)$^*$ & 19(3) & - & $E2$ & & &\\	
              & $2^+$ & $4^+_1$ & 1271.1(8)$^*$ & 7.6(11) & & $(E2)$& & &\\
              & $2^+$ & $3^-_1$ & 1165.3(2) & 8.3(13) & & $(E1)$ & & &\\
              & $2^+$ & $2^+_3$ & 1042.8(8)$^*$ & 17(2) & & & & &\\
              & $2^+$ & $2^+$ & 797.6(8)$^*$ & 8.6(13) & & & & &\\
              & $2^+$ & $4^+_3$ & 735.1(5)$^*$ & 1.5(5) & & $(E2)$ & & &\\
\hline
3524.0(4) & $2^+$ & $2^+_1$ & 2267.7(2) & 100 & -0.07(40) & $M1+E2$ &  0.08(2) & 1100(400) & - \\	 	
          & $2^+$ & $4^+_1$ & 1276.8(4) & 37(6) & & $(E2)$ & & &\\
          & $2^+$ & $3^-_1$ & 1169.5(9)$^*$ & 11(2) & & $(E1)$ & & &\\          
          & $2^+$ & $3^+_1$ & 768.3(2)$^*$ & 29(4) & & & & &\\
          & $2^+$ & $< 5$ & 759.4(4)$^*$ & 29(7) & & & & &\\
          & $2^+$ & $0^+$ & 538.7(3)$^*$ & 5(2) & - & $E2$ & & &\\
          & $2^+$ & $2^+$ & 431.5(4) & 5(2) & & & & &\\
\hline
3526.5(2)$^{\dagger}$ & $(1,2^+)$ & $0^+$ & 3526.5(2)$^*$ & 100 & & &  0.25(2) & 230(30) & $>$\,180 \\ 	
          & $(1,2^+)$ & $2^+_2$ & 1375.6(2)$^*$ & 40(6) & & & & &\\
          & $(1,2^+)$ & $2^+$ & 805.9(4)$^*$ & 10(2) & & & & &\\
\hline
3529.2(3) & $(4^+)$ & $2^+_2$ & 1378.7(2) & 100 & & & & & \\	
          & $(4^+)$ & $4^+_1$ & 1282.5(3) & 87$^{+13}_{-15}$ & & & & & \\
          & $(4^+)$ & $4^+_2$ & 1008.8(2) & 43(9) & & & & & \\
          & $(4^+)$ & $6^+_1$ & 980.1(3)$^*$ & 41(9) & & & & & \\
          & $(4^+)$ & $4^+$ & 380.5(4) & 24(5) & & & & & \\
\hline
3553.2(2) & $(3)^-$ & $2^+_1$ & 2296.8(2) & 100 & & &  0.187(14) & 460(130) & 240$^{+\,160}_{-\,80}$ \\
          & $(3)^-$ & $3^+_1$ & 797.7(3)$^*$ & 20(4) & & & & &\\
\hline
3557.8(2) & $(4^+)$ & $3^-_1$ & 1204.1(2) & 100 & & & & & \\
\hline
3583.2(4) & $(2^+,4^+)$ & $2^+_1$ &	2326.9(2)$^*$ & 23(6) & & & 0.11(5) & 770(540) & - \\ 	
		  & $(2^+,4^+)$ & $2^+_2$ & 1433.2(5)$^*$ & 13(7) & & & & & \\ 
		  & $(2^+,4^+)$ & $3^-_1$ & 1229.0(5)$^*$ & 19(11) & & & & & \\ 
		  & $(2^+,4^+)$ & $2^+_3$ & 1107.8(2)$^*$ & 100 & & & & & \\ 
		  & $(2^+,4^+)$ & $4^+$ & 669.8(5)$^*$ & 72(13) & & & & & \\ 
		  & $(2^+,4^+)$ & $2^+$ & 617.1(4)$^*$ & 21(5) & & & & & \\ 
\hline
3601.6(2)$^{\dagger}$ & $2^+$ & $0^+_1$ & 3601.6(2)$^*$ & 100 & - & $E2$ & 0.11(2) & 730(200) & - \\	
		                & $2^+$ & $2^+_1$ & 2345.5(3)$^*$ & 63(10) & & & & & \\ 
		                & $2^+$ & $0^+_2$ & 1411.4(2)$^*$ & 29(8) & - & $E2$ & & & \\ 
		                & $2^+$ & $4^+_2$ & 1081.8(2)$^*$ & 29(5) & & $(E2)$ & & & \\ 
		                & $2^+$ & $2^+$ & 881.1(3)$^*$ & 15(3) & & & & & \\ 
		                & $2^+$ & $4^+$ & 452.9(4)$^*$ & 10(2) & & $(E2)$ & & & \\ 
\hline
3603.1(2) & $\leq 6$ & $4^+_1$ & 1356.1(3) & 100 & & & & & \\
\hline
3610.8(4) & $(2^+,3,4^+)$ & $2^+_1$ & 2354.2(2) & 100 & & & 0.44(7) & 90(30) & 111$^{+60}_{-34}$ \\
          & $(2^+,3,4^+)$ & $2^+_2$ & 1459.9(5) & 73(12) & & & & & \\
          & $(2^+,3,4^+)$ & $4^+_1$ & 1364.2(6)$^*$ & 15(6) & & & & & \\
\hline
3643.4(3)$^{\dagger}$ & $(2^+,3,4^+)$ & $4^+_1$ & 1396.4(2)$^*$ & 36(7) & & & & & \\
          & $(2^+,3,4^+)$ & $4^+_2$ & 1122.9(2)$^*$ & 73(12) & & & & & \\
          & $(2^+,3,4^+)$ & $2^+$ & 922.8(3)$^*$ & 65(12) & & & & & \\
          & $(2^+,3,4^+)$ & $4^+$ & 726.4(2)$^*$ & 100 & & & & & \\
\hline
3654.1(2) & $2^+$ & $0^+_1$ & 3654.1(2) & 100 & - & $E2$ & 0.36(3) & \multirow{2}{*}{140(20)} & \multirow{2}{*}{-} \\ 	
		  &	$2^+$ & $2^+_1$ & 2397.8(2) & 79(12) & & & 0.31(4) & & \\ 	
\hline
3688.0(6)$^{\dagger}$ & $(1,2^+)$ & $0^+_1$ & 3688.3(3)$^*$ & 60(11) & - & & & & \\
          & $(1,2^+)$ & $2^+_1$ & 2431.5(4)$^*$ & 65(12) & & & & & \\
          & $(1,2^+)$ & $2^+_2$ & 1537.5(3)$^*$ & 40(8) & & & & & \\
          & $(1,2^+)$ & $2^+$ & 721.8(2)$^*$ & 100 & & & & & \\
\hline
3705.6(5)$^{\dagger}$ & $(2^+,3,4^+)$ & $2^+_2$ & 1554.8(3)$^*$ & 52(8) & & & & & \\
          & $(2^+,3,4^+)$ & $4^+_1$ & 1459.0(3)$^*$ & 100 & & & & & \\
\hline
3719.6(2)$^{\dagger}$ & $(2^+,3,4^+)$ & $2^+_1$ & 2462.9(2)$^*$ & 100 & & & & & \\
          & $(2^+,3,4^+)$ & $4^+_1$ & 1472.7(3)$^*$ & 24(6) & & & & & \\
\hline
3725.3(2) & $(1,2^+)$ & $2^+_1$ & 2468.8(2) & 100 & - & & & & \\
          & $(1,2^+)$ & $2^+_2$ & 1574.7(3)$^*$ & 30(6) & & & & & \\
          & $(1,2^+)$ & $0^+_2$ & 1534.8(3)$^*$ & 61(15) & - & & & & \\
\hline
3774.5(4)$^{\dagger}$ & $2^+$ & $0^+_1$ & 3774.3(5)$^*$ & 39(8) & - & $E2$ & & & \\
          & $2^+$ & $2^+_1$ & 2518.0(2)$^*$ & 100 & & & & & \\
          & $2^+$ & $4^+_1$ & 1527.3(2)$^*$ & 56(10) & & $(E2)$ & & & \\
          & $2^+$ & $3^-_1$ & 1420.6(3)$^*$ & 68(12) & & $(E1)$ & & & \\
          & $2^+$ & $2^+$ & 808.8(8)$^*$ & 29(7) & & & & & \\
\hline
3781.0(5) & $(2^+,3,4^+)$ & $2^+_2$ & 1630.1(2) & 100 & & & & & \\
          & $(2^+,3,4^+)$ & $4^+$ & 997.8(7)$^*$ & 54(11) & & & & & \\
\hline
3827.1(3) & $(1^-,2^+)$ & $0^+_1$ & 3827.1(2)$^*$ & 100 & - & & 0.41(2) & \multirow{2}{*}{108(9)} & \multirow{2}{*}{-} \\ 	
		      & $(1^-,2^+)$ & $2^+_1$ & 2570.8(2)$^*$ & 50(8) & & & 0.26(4) & & \\ 	
		      & $(1^-,2^+)$ & $3^-_1$ & 1473.0(7)$^*$ & 23(5) & & & & & \\
\hline
3873.4(3) & $(1,2^+)$ & $0^+_1$ & 3873.4(3)$^*$ & 100 & - & & 0.95(4) & 5(3) & - \\	
\hline
3913.5(2)     & $ 2^+ $ & $0^+_1$ & 3913.5(2)$^*$ & 100 & - & $E2$ & 0.38(3) & 120(20) & - \\	
		      & $ 2^+ $ & $4^+_2$ & 1392.4(3)$^*$ & 23(4) & & $(E2)$ & & & \\
\hline
3925.5(8)$^{\dagger}$  & $(1,2^+)$ & $0^+_1$ & 3926.0(6)$^*$ & 38(8) & - & & & & \\	
		               & $(1,2^+)$ & $2^+_1$ & 2668.4(10)$^*$ & 100 & & & 0.16(2) & 410(70)& - \\
\hline
3984.7(3)     & $(1^-,2^+)$ & $0^+_1$ &	3984.7(3)$^*$ & 100 & - & & 0.53(5) & 64(12) & - \\ 
		      & $(1^-,2^+)$ & $3^-_1$ & 1630.0(3)$^*$ & 17(3) & & & & & \\
		      & $(1^-,2^+)$ & $2^+_3$ & 1507.8(4)$^*$ & 9(2) & & & & & \\
\hline
4019.1(9)$^{\dagger}$ & $(1,2^+)$ & $0^+_1$ & 4018.4(6)$^*$ & 100 & - & & 0.54(6)& 60(20) & - \\	
		              & $(1,2^+)$ & $0^+_2$ & 1828.9(6)$^*$ & 73(15) & - & & & &  \\
\hline
4044.0(2)$^{\dagger}$ & $(1,2^+)$ & $0^+_1$ & 4044.2(8)$^*$ & 21(5) & - & & & &  \\	
		              & $(1,2^+)$ & $2^+_1$ & 2787.5(5)$^*$ & 100 & & & & &  \\
		              & $(1,2^+)$ & $2^+_2$ & 1893.3(4)$^*$ & 78(13) & & & 0.18(4) & 350(110) & - \\
\hline
4077.2(10)$^{\dagger}$ & $(1,2^+)$ & $0^+_1$ & 4076.6(5)$^*$ & 87$^{+13}_{-17}$ & - & & & &  \\	
		               & $(1,2^+)$ & $2^+_1$ & 2819.6(10)$^*$ & 100 & & & & &  \\
		               & $(1,2^+)$ & $2^+_2$ & 1927.0(3)$^*$ & 63(14) & & & & &  \\
		               & $(1,2^+)$ & $2^+_3$ & 1602.9(12)$^*$ & 40(10) & & & & &  \\
\hline
4086.5(2)$^{\dagger}$ & $(1,2^+)$ & $0^+_1$ & 4086.3(4)$^*$ & 100 & - & & & &  \\	
		              & $(1,2^+)$ & $2^+_1$ & 2829.9(5)$^*$ & 78(19) & & & & &  \\
		              & $(1,2^+)$ & $2^+_2$ & 1936.1(3)$^*$ & 48(14) & & & & &  \\
\hline
4096.7(2) & $< 5$ & $2^+_1$ & 2840.2(2)$^*$ & 100 & & & & &  \\	
\hline
4141.2(3) & $1^-$ & $0^+_1$ & 4141.2(3) & 100 & - & $E1$ & 0.55(2) & 59(5) & 39(9)$^{\mathrm{d}}$  \\	
\hline
4160.5(3) & $1^-$ & $0^+_1$ & 4160.5(3) & 100 & - &$E1$ & 0.77(5) & 23(6) & 14.9(14)$^{\mathrm{d}}$ \\
\end{longtable*}

\renewcommand*{\arraystretch}{1.2}
\begin{longtable*}{p{0.10\textwidth}p{0.09\textwidth}p{0.07\textwidth}p{0.10\textwidth}p{0.07\textwidth}p{0.12\textwidth}p{.10\linewidth}p{0.09\textwidth}p{0.09\textwidth}p{0.08\textwidth}}
\caption{Experimental data for excited states in $^{114}$Sn. See Table\,\ref{tab:112Sn_tau} for more information. The multipole-mixing ratios $\delta$ correspond to the adopted values\,\cite{ENSDF}. Only statistical uncertainties are given for the lifetime values. As explained in the text, systematic uncertainties should be considered at the 19\,$\%$ level.}
\label{tab:114Sn_tau}

\vspace*{1mm}
\\
\toprule
$E_x$\,[keV]& $J^{\pi}_i$ & $J^{\pi}_f$ & $E_{\gamma}$\,[keV] & $I_{\gamma}$\,[$\%$] & $\delta$ & $\Pi L$ & $F(\tau)$ & $\tau$\,[fs] & $\tau_{\mathrm{lit.}}$\,[fs]\\
\hline

\endfirsthead

\multicolumn{7}{c}{{\small Table \ref{tab:114Sn_tau}: Continuation}} \\

\hline
\hline

$E_x$\,[keV]& $J^{\pi}_i$ & $J^{\pi}_f$ & $E_{\gamma}$\,[keV] & $I_{\gamma}$\,[$\%$] & $\delta$ & $\Pi L$ & $F(\tau)$ & $\tau$\,[fs] & $\tau_{\mathrm{lit.}}$\,[fs]\\
\hline
\endhead
\endfoot
\hline
\hline
\multicolumn{10}{l}{$^{\mathrm{a}}$ Taken from Ref.\,\cite{ENSDF}.}\\
\multicolumn{10}{l}{$^{\mathrm{b}}$ Taken from Ref.\,\cite{Gabl01a}.}\\
\multicolumn{10}{l}{$^{\#}$ No clear assignment possible.}
\endlastfoot

1299.7(2)& $2^+_1$ & $0^+_1$ & 1299.7(2)& 100 & -& $E2$ & 0.145(13) & 590(70) & 610(40)$^{\mathrm{a}}$\\
\hline	
1952.9(2)& $0^+_2$ & $2^+_1$ & 653.2(2)& 100 & - & $E2$ & & & 9(3)\,ps$^{\mathrm{a}}$\\
\hline
2155.9(2)& $0^+_3$ & $2^+_1$ & 856.2(2)& 100 & -& $E2$ & & & $> 11$\,ps$^{\mathrm{a}}$\\
\hline
2187.3(3)& $4^+_1$ & $2^+_1$ & 887.6(2)& 100 & & $(E2)$ & & & 7.6(6)\,ps$^{\mathrm{a}}$\\
\hline
2238.6(2)& $2^+_2$ & $0^+_1$ & 2238.5(2) & 100 & - & $E2$ & $< 0.04$  & $> 2100$ & - \\
         & $2^+_2$ & $2^+_1$ & 938.9(2) & 81(12) & -7.1$^{+1.2}_{-1.9}$ & $M1+E2$ &  &  &  \\
         & $2^+_2$ & $0^+_2$ & 286.5(10) & 0.9(3) & - & $E2$ &  &  &  \\
\hline
2274.5(2)& $3^-_1$ & $2^+_1$ & 974.8(2) & 100 &  & $(E1)$ & 0.123(11)  & 700(80) & 520(30)$^{\mathrm{a}}$ \\
\hline
2420.5(2)& $0^+_4$ & $2^+_1$ & 1120.8(2)& 100 & - & $E2$ & & & \\
\hline
2453.8(2)& $2^+_3$ & $0^+_1$ & 2453.7(2) & 28(4) & - & $E2$ & 0.04(2)  & \multirow{2}{*}{2700(1100)} & \multirow{2}{*}{-} \\
         & $2^+_3$ & $2^+_1$ & 1154.0(2) & 100 & -2.8$^{+1.8}_{-9.5}$ & $M1+E2$ & 0.05(2)  &  &  \\
         & $2^+_3$ & $2^+_2$ & 215.4(4) & 1.3(3) & & & & &  \\
\hline
2514.4(2)& $3^+_1$ & $4^+_1$ & 327.1(2) & 100 & 0.02$^{+0.02}_{-0.01}$ & $M1+E2$ & & & \\
\hline
2613.7(4)& $4^+_2$ & $2^+_1$ & 1314.5(2) & 100 &  & $(E2)$ & 0.100(9) & 920(130) & 793(144)$^{\mathrm{b}}$ \\
         & $4^+_2$ & $4^+_1$ & 426.0(4) & 1.6(6) & -0.24$^{+0.06}_{-0.05}$ & $M1+E2$ & & & \\
         & $4^+_2$ & $2^+_2$ & 375.2(3) & 1.8(6) &  & $(E2)$ & & & \\
\hline
2764.9(5)& $4^+$ & $2^+_1$ & 1465.3(2) & 100 &  & $(E2)$ & 0.042(19) & 2900(1600) & 808(433)$^{\mathrm{b}}$ \\
         & $4^+$ & $4^+_1$ & 577.3(3)$^*$ & 2.3(9) & & & & & \\
         & $4^+$ & $2^+_2$ & 525.7(2) & 1.1(6) & & $(E2)$ & & & \\
         & $4^+$ & $3^-_1$ & 490.3(3) & 1.4(7) & & $(E1)$ & & & \\
         & $4^+$ & $3^+_1$ & 251.1(3) & 4.3(10) & $-0.1^{+0.1}_{-4.2}$& $M1+E2$ & & & \\
\hline
2814.6(2)& $5^-_1$ & $4^+_1$ & 627.4(2) & 100 &  & $(E1)$ &  & & $> 2020$\\
         & $5^-_1$ & $3^-_1$ & 539.9(2) & 13(3) &  & $(E2)$ & & & \\
\hline
2859.2(5)& $4^+$ & $2^+_1$ & 1559.7(2) & 100 &  & $(E2)$ & 0.104(10) & 900(130) & - \\
         & $4^+$ & $4^+_1$ & 672.1(4)$^*$ & 4.2(12) & & & & & \\
         & $4^+$ & $2^+_2$ & 619.8(3)$^*$ & $< 1.5$ & & $(E2)$ & & & \\
         & $4^+$ & $2^+_3$ & 405.5(3)$^*$ & $< 1.7$ & & $(E2)$ & & & \\
\hline
2904.9(3)& $3^-$ & $2^+_1$ & 1605.1(4) & 3.4(7) & & $(E1)$ & & & \\
         & $3^-$ & $4^+_1$ & 717.3(2) & 100 & $-0.7^{+0.2}_{-0.4}$ & $(E1)$ & 0.098(25) & 880(360) & - \\
         & $3^-$ & $3^+_1$ & 390.2(2) & 26(4) & & & & & \\
         & $3^-$ & $4^+_2$ & 290.3(4) & 1.4(5) & & $(E1)$ & & & \\
\hline
2915.6(2)& $2^+$ & $0^+_1$ & 2915.5(2) & 100 & - & $E2$ & 0.067(6)  & \multirow{2}{*}{1600(200)} & \multirow{2}{*}{-} \\
         & $2^+$ & $2^+_1$ & 1615.8(2) & 29(4) & $0.08 < \delta < 1.7$ & $M1+E2$ & 0.06(2)  &  &  \\
\hline
2943.4(2)& $2^+$ & $0^+_1$ & 2943.4(6) & 2.5(7) & - & $E2$ &  & &\\
         & \multirow{2}{*}{$2^+$} & \multirow{2}{*}{$2^+_1$} & \multirow{2}{*}{1643.3(2)} & \multirow{2}{*}{100} & -0.61(15) & $M1+E2$ & \multirow{2}{*}{$< 0.04$}  & \multirow{2}{*}{$> 3200$} & \multirow{2}{*}{-} \\
         &  &   &   &  & -7$^{+10}_{-3}$ & $M1+E2$ & & &  \\
         & $2^+$ & $0^+_2$ & 990.3(3) & 7.7(13) & - & $E2$ &  & &\\
         & $2^+$ & $2^+_2$ & 704.3(3) & 3.2(7) & & &  & &\\
         & $2^+$ & $3^-_1$ & 668.3(2) & 87(12) & & $(E1)$ &  & &\\
         & $2^+$ & $0^+_4$ & 522.4(5) & 1.0(5) & - & $E2$ & & &\\
         & $2^+$ & $2^+_3$ & 489.6(2) & 2.9(6) &  & &  & &\\
\hline
3024.9(2)& $2^+$ & $2^+_1$ & 1725.4(2) & 100 & & & & & \\
         & $2^+$ & $0^+_2$ & 1071.7(4)$^*$ & 9(2) & - & $E2$ & & & \\
         & $2^+$ & $2^+_2$ & 786.4(2)$^*$ & 13(2) & & & & & \\
         & $2^+$ & $2^+_3$ & 571.1(2)$^*$ & 12(2) & & & & & \\
\hline
3028.1(2)& $0^+$ & $2^+_1$ & 1728.4(2) & 100 & - & $E2$ & 0.10(3) & 900(400) & -\\
         & $0^+$ & $2^+_2$ & 789.4(5)$^*$ & 1.6(10) & - & $E2$ & & & \\
         & $0^+$ & $2^+_3$ & 574.1(3)$^*$ & 3.8(9) & - & $E2$ & & & \\
\hline
3185.5(2)& $2^+$ & $0^+_1$ & 3185.5(2) & 58(8) & - & $E2$ & 0.25(2) & \multirow{3}{*}{209(17)} & \multirow{3}{*}{-}\\
         & \multirow{2}{*}{$2^+$} & \multirow{2}{*}{$2^+_1$} & \multirow{2}{*}{1885.8(2)} & \multirow{2}{*}{100} & -0.27(7) & $M1+E2$ & \multirow{2}{*}{0.28(2)} &  &\\
         & & & & & 7$^{+5}_{-2}$ & $M1+E2$ & & &\\         
\hline
3206.6(6)& $4^+$ & $2^+_1$ & 1907.2(3) & 100 & & $(E2)$& & & \\
         & $4^+$ & $4^+_1$ & 1019.6(4) & 29(7) & & & & & \\
         & $4^+$ & $3^-_1$ & 932.3(2) & 8(3) & & $(E1)$& & & \\
         & $4^+$ & $4^+_2$ & 592.0(9) & 7(3) & & & & & \\
\hline
3211.3(2)& $(1,2^+)$ & $0^+_1$ & 3211.2(2) & 100 & & & 0.16(3) & \multirow{4}{*}{390(80)} & \multirow{4}{*}{-}\\
         & $(1,2^+)$ & $2^+_1$ & 1911.8(2) & 38(6) & & & 0.14(4) & & \\
         & $(1,2^+)$ & $0^+_2$ & 1258.0(7)$^*$ & 24(4) & & & & & \\
         & $(1,2^+)$ & $0^+_3$ & 1054.6(2)$^*$ & 42(6) & & & 0.21(6) & & \\
         & $(1,2^+)$ & $2^+_3$ & 757.5(2)$^*$ & $< 6$ & & & & & \\
\hline
3225.1(2)& $3^-$ & $2^+_1$ & 1925.4(2) & 100 &  & $(E1)$ & 0.15(2) & 500(95) & - \\
         & $3^-$ & $2^+_3$ & 771.4(4) & 2.1(8) & & $(E1)$ & & & \\
         & $3^-$ & $3^-$ & 319.9(4) & 6.6(14) & & & & & \\
\hline
3308.5(2)& $0^+$ & $2^+_1$ & 2008.8(2) & 100 & - & $E2$ & & & \\
\hline
3326.4(2)& $2^+$ & $0^+_1$ & 3326.2(2)$^*$ & 59(9) & - & $E2$ & 0.11(4) & 750(380)& - \\
         & $2^+$ & $2^+_1$ & 2026.4(2)$^*$ & 100 & & & & & \\
         & $2^+$ & $0^+_2$ & 1373.2(3)$^*$ & 24(4) & - & $E2$ & & & \\
         & $2^+$ & $4^+_1$ & 1139.0(3)$^*$ & 13(2) & & $(E2)$ & & & \\
\hline
3356.3(7)& $4^+$ & $2^+_1$ & 2057.1(2) & 100 & & $(E2)$ & & & \\
         & $4^+$ & $4^+_1$ & 1168.6(6)$^*$ & 6(3) & & & & & \\
\hline
3397.3(2)& $3^-$ & $2^+_1$ & 2097.6(2)$^*$ & 72(11) &  & & 0.24(6) & 250(90) & - \\
         & $3^-$ & $2^+_2$ & 1158.3(2)$^*$ & 30(5) & & & & & \\
         & $3^-$ & $3^-_1$ & 1122.0(4) & 100 & -0.4$^{+0.2}_{-0.7}$ & $M1+E2$ & & & \\
         & $3^-$ & $2^+_3$ & 943.2(2)$^*$ & 27(4) & & & & & \\
\hline
3422.0(3)& $0^+$ & $2^+_1$ & 2121.9(4) & 100 & - & $E2$ & & & \\
         & $0^+$ & $2^+_2$ & 1182.9(5)$^*$ & 41(10) & - & $E2$ & & & \\
         & $0^+$ & $2^+_3$ & 968.2(4)$^*$ & 21(7) & - & $E2$ & & & \\
\hline
3452.1(2)& $(1^-)$ & $0^+_1$ & 3452.1(2) & 100 &  & $(E1)$ & 0.93(3) & 6(3) & - \\
\hline
3478.1(4)& $2^+$ & $0^+_1$ & 3478.1(4) & 10(2) & - & $E2$ & & & \\
         & $2^+$ & $2^+_1$ & 2178.5(2) & 100 & & & & & \\
         & $2^+$ & $0^+_2$ & 1524.4(3) & 13(3) & - & $E2$ & & & \\
         & $2^+$ & $2^+_2$ & 1240.0(2) & 83(13) & & & & & \\
         & $2^+$ & $3^-_1$ & 1203.3(2) & 67(10) & & $(E1)$ & & & \\
         & $2^+$ & $2^+_3$ & 1025.1(2)$^*$ & 21(4) & & & & & \\
         & $2^+$ & $3^+_1$ & 962.9(3) & 52(8) & & & & & \\
         & $2^+$ & $4^+$   & 619.7(4) & 11(3) & & $(E2)$ & & & \\
         & $2^+$ & $3^-$ & 572.4(4) & 14(3) & & & & & \\
\hline
3483.9(4)$^{\dagger}$& $1^-,2^+$ & $2^+_1$ & 2184.1(2) & 100 & & & 0.16(3) & \multirow{3}{*}{450(110)} & \multirow{3}{*}{-}  \\
         & $1^-,2^+$ & $0^+_3$ & 1327.7(3)$^{\#}$ & 14(2) & & & & & \\
         & $1^-,2^+$ & $3^-_1$ & 1209.0(2) & 35(2) & & & 0.10(4) & & \\
\hline
3487.5(4)& $5^-$ & $3^-_1$ & 1213.3(4)$^*$ & 40(13) &  & $(E2)$ & & & \\
         & $5^-$ & $3^-$ & 582.3(2)$^*$ & 100 &  & $(E2)$ & & & \\
\hline
3494.3(3)$^{\dagger}$& $1,2^+$ & $0^+_1$ & 3494.2(3) & 81(13) & & & 0.18(5) & \multirow{2}{*}{450(130)} & \multirow{2}{*}{-} \\
         & $1,2^+$ & $2^+_1$ & 2194.4(2) & 100 & & & 0.12(3) & & \\
         & $1,2^+$ & $0^+_2$ & 1540.5(3) & 26(5) & & & & & \\
\hline
3514.1(3)& $3^-$ & $2^+_1$ & 2214.4(2) & 100& & $(E1)$ & 0.27(8) & 206(93) & - \\
         & $3^-$ & $4^+_1$ & 1327.0(4)$^{\#}$ & 9(3) & &  $(E1)$ & & & \\
         & $3^-$ & $2^+_2$ & 1275.0(2)$^*$ & 23(4) & & $(E1)$ & & & \\         
\hline
3524.4(2)& $3^-$ & $2^+_1$ & 2224.5(3)$^*$ & 100 & & $(E1)$ & $< 0.09$ & & \\
         & $3^-$ & $4^+_1$ & 1337.0(2) & 27(5) & &  $(E1)$ & & & \\
         & $3^-$ & $3^-_1$ & 1249.6(3) & 24(4) & & $(E2)$ & 0.10(5) & 900(690) & - \\         
         & $3^-$ & $3^+_1$ & 1010.1(3) & 31(6) &  & & & & \\
\hline
3547.6(2)& $0^+$ & $2^+_2$ & 1308.9(2)$^*$ & 92$^{+8}_{-21}$ & - & $E2$ & & & \\
         & $0^+$ & $2^+_3$ & 1093.8(3)$^*$ & 100 & - & $E2$ & & & \\
\hline
3560.8(2)& $2^+$ & $0^+_1$ & 3560.8(2) & 100 & - & $E2$ & 0.13(3) & 590(190)& - \\
         & $2^+$ & $2^+_1$ & 2261.1(3)$^*$ & 19(4) & & & & & \\
         & $2^+$ & $0^+_3$ & 1404.9(4) & 14(3) & - & $E2$ & & & \\
\hline
3610.2(4)& $5^{(-)}$ & $4^+_1$ & 1422.9(3) & 100 &  & $(E1)$ & 0.33(6) & 150(40) & - \\
\hline
3650.3(3)$^{\dagger}$& $1^-,2^+$ & $0^+_1$ & 3650.1(3) & 100 & & & 0.20(5) & 320(120)& - \\
         & $1^-,2^+$ & $2^+_1$ & 2350.3(3) & 26(5) & & & & & \\
         & $1^-,2^+$ & $0^+_3$ & 1493.7(3) & 21(4) & & & & & \\
         & $1^-,2^+$ & $3^-_1$ & 1374.6(2) & 82(13) & & & & & \\
\hline
3679.5(4)$^{\dagger}$& $1,2^+$ & $0^+_1$ & 3679.4(2) & 100 & & & 0.30(4) & \multirow{2}{*}{140(20)} & \multirow{2}{*}{-} \\
         & $1,2^+$ & $2^+_1$ & 2379.5(3) & 81(12) & & & 0.37(5) & & \\
\hline
3692.5(3)& $2^+$ & $0^+_1$ & 3692.8(3)$^*$ & 100 & - & $E2$ & 0.13(4) & 580(250) & \\
         & $2^+$ & $2^+_1$ & 2392.3(3)$^*$ & 47(8) & & & & & \\
\hline
3722.5(3)& $(2^+)$ & $2^+_1$ & 2422.5(2)$^*$ & 73(12) & & & 0.30(7) & 180(60) & \\
         & $(2^+)$ & $3^-_1$ & 1446.6(2) & 100 & & & & & \\
\hline
3792.2(3)& $1,2^+$ & $0^+_1$ & 3792.2(3)$^*$ & 100 & & & & & \\
         & $1,2^+$ & $2^+_1$ & 2492.7(5)$^*$ & 47(13) & & & & & \\
\hline
3869.4(5)$^{\dagger}$& $2^+$ & $0^+_1$ & 3869.2(5) & 100 & - & $E2$ & 0.33(8) & \multirow{4}{*}{120(30)} & \\
         & $2^+$ & $2^+_1$ & 2569.1(11) & 19(9) & & & & & \\
         & $2^+$ & $4^+_1$ & 1682.7(4) & 55(12) & & $E2$ & 0.40(6) & & \\
         & $2^+$ & $3^-_1$ & 1595.2(2) & 18(6) & & $(E1)$ & & & \\
\hline
3933.0(4)& $1,2^+$ & $0^+_1$ & 3933.0(4)$^*$ & 100 & & & 0.79(8) & 19(9) & - \\
         & $1,2^+$ & $2^+_1$ & 2633.6(3)$^*$ & 24(5) & & & & & \\
\hline
4022.4(3)$^{\dagger}$& $1,2^+$ & $0^+_1$ & 4022.4(3) & 100 & & & 0.81(4) & 18(5) & - \\
\end{longtable*}

\subsection{Lifetimes of the $2^+_1$ and $3^-_1$ states}
\label{sec:tau_comp}

In general, the lifetimes determined in this work are in excellent agreement with lifetimes reported in Refs.\,\cite{Kum05a,Orce07a,Jungcl11a,Gabl01a} and also the lower limits found are in good agreement with previously known lifetimes, see, {\it e.g.}, $\tau \left( 2^+_2 \right)$ and $\tau \left( 4^+_1 \right)$ in Table~\ref{tab:112Sn_tau}.
\\
\\ 

{\it $J^{\pi} = 2^+_1$:} A rather obvious inconsistency between lifetime measurements employing Doppler-shift methods\,\cite{Jungcl11a} and from Coulex experiments\,\cite{Allm15a, Kum17a} has been observed for $\tau \left( 2^+_1 \right)$ in the stable even-even Sn isotopes and is apparently also seen for the unstable tin isotopes, see the recent work on $^{110}$Sn\,\cite{Kumb16a}. Our present $(p,p'\gamma)$ experiments might support the lifetimes, which have been reported in Ref.\,\cite{Jungcl11a} by A.\,Jungclaus {\it et al.}, see Tables\,\ref{tab:112Sn_tau} and \ref{tab:114Sn_tau}. It should be noted that due to the kinematics of the $(p,p'\gamma)$ reaction, our systematic uncertainties are dominated by the variation of the nuclear stopping power in contrast to Ref.\,\cite{Jungcl11a} where the electronic stopping dominates the systematic uncertainties. J.N. Orce {\it et al.} first reported a $\tau \left( 2^+_1 \right)$ of $750^{+125}_{-90}$\,fs in $^{112}$Sn using INS-DSAM and later revised their measured lifetime to $530^{+100}_{-80}$\,fs\,\cite{Orce07a}, i.e. closer to the presently adopted value. The authors argued that one should introduce a correction to the recoil-velocity distribution when using neutrons with an energy ``well above'' the excitation threshold. In the light of the new data, this discussion might not be necessary since the inital recoil velocity can be determined precisely from the $p\gamma$ coincidence data and since feeding can be excluded due to the excitation-energy gate. We want to mention that $\tau \left( 2^+_1 \right)$ in $^{114}$Sn was determined using the $^{112}$Sn data set. Here, the $^{114}$Sn admixture to the target accounted to roughly 13\,$\%$. Due to a large $^{116}$Sn admixture ($\sim 10\,\%$) in the $^{114}$Sn target, it was not possible to unambiguously determine the energy-centroid shifts of the two close-lying $2^+$ states of $^{114}$Sn and $^{116}$Sn.
\\
\\
Since some Coulex experiments and especially those with radioactive ion beams, see, {\it e.g.}, Refs.\,\cite{Jungcl11a, Doorn08a, Bad13a}, rely on the normalization to ``well-known'' $B(E2)$ values in stable nuclei, it is important to resolve these discrepancies. The lifetime of the $2^+_1$ in $^{116}$Sn should be certainly remeasured as well.
\\ 

{\it $J^{\pi} = 3^-_1$:} The lifetime $\tau \left( 3^-_1 \right) = 280(20)$\,fs in $^{112}$Sn agrees nicely with the previously reported value of A.\,Jungclaus {\it et al.}\,\cite{Jungcl11a}, while it is in conflict with the one reported by A.\,Kumar {\it et al.} of $510^{+200}_{-120}$\,fs\,\cite{Kum05a}. Possibly, the latter discrepancy might be attributed to feeding missed in the $(n,n'\gamma)$ experiment. The lifetime $\tau \left( 3^-_1 \right) = 700(80)$\,fs in $^{114}$Sn does, however, not confirm the value of 520(30)\,fs measured by A.\,Jungclaus {\it et al.}\,\cite{Jungcl11a}. It should be mentioned that also $\tau \left( 3^-_1 \right)$ could be estimated from the $^{112}$Sn data set for $^{114}$Sn. Both values measured in the $(p,p'\gamma)$ experiments are consistent.

\subsection{$\gamma$-decay intensities, newly observed and non-observed $\gamma$ decays}

The total photopeak efficiency of the combined SONIC@HORUS setup was already shown in Fig.\,\ref{fig:efficiency}. An uncertainty of less than 10\,$\%$ due to the geometry of the $^{56}$Co source is included in the uncertainties given for the $\gamma$-decay intensities $I_{\gamma}$ in Tables~\ref{tab:112Sn_tau} and \ref{tab:114Sn_tau}. If $\gamma$-decay branching of an excited state was observed, the respective $\gamma$-decay branching ratio could be calculated as follows:

\begin{equation*}
I_{\gamma} = \frac{A_i \varepsilon(E_{\gamma,j})}{A_j \varepsilon(E_{\gamma,i})}
\end{equation*}

where $A_i$ is the peak intensity of a $\gamma$ decay with decay energy $E_{\gamma,i}$ corrected by the corresponding detection efficiency $\varepsilon(E_{\gamma,i})$. Usually, one should also correct for the $p\gamma$-angular correlation, detector deadtimes, and the system deadtime. This correction was previously estimated to be less than 20\,$\%$\,\cite{Henn14a}. 
\\

A comparison of the $\gamma$-decay branching ratios determined in this work with adopted ratios\,\cite{ENSDF} showed that this statement is in general correct. Within the statistical uncertainties very good agreement was obtained. For instance, for the $2^+_2$ state of $^{114}$Sn at 2238.6(2)\,keV the following $\gamma$-decay intensities are adopted\,\cite{ENSDF}: 100\,$\%$ ($0^+_1$), 82(2)\,$\%$ ($2^+_1$) and 0.8(3)\,$\%$ ($0^+_2$) which are in perfect agreement with our results, see Table~\ref{tab:114Sn_tau}. The same holds for the $2^+_3$ state of $^{114}$Sn and the $2^+_2$ state of $^{112}$Sn, compare Table~\ref{tab:112Sn_tau}, as well as for states with $J^{\pi} \neq 2^+$, {\it e.g.}, the $4^+$ state of $^{114}$Sn at 2764.9(5)\,keV or the $3^-$ state at 2904.9(3)\,keV. For these states the very weak $\gamma$-decays with intensities of about 1\,$\%$ were also observed in our experiment. However, discrepancies are already observed for the $2^+_3$ state of $^{112}$Sn where a $\gamma$-decay intensity of 20(2)\,$\%$ to the $2^+_1$ state was previously reported\,\cite{ENSDF,Wig76a,Kum05a}. In our experiment an intensity of 36(5)$\%$ was observed. We cannot comment on the efficiency calibration of Ref.\,\cite{Wig76a}, however, Ref.\,\cite{Kum05a} used a $^{226}$Ra source for the efficiency calibration which only provides a reliable efficiency calibration up to 2.45\,MeV. Apparently, the efficiency has been underestimated since discrepancies for the $\gamma$-decay intensities are observed for decays with $E_{\gamma} > 2.45$\,MeV. For instance, Ref.\,\cite{Wig76a} reported an $I_{\gamma}$ of 15.9(13)\,$\%$ for the decay of the $2^+$ state of $^{112}$Sn at 2720.6(2)\,keV to the ground state while Ref.\,\cite{Kum05a} gave a value of 33(6)\,$\%$. Our analysis provides a value of 10.0(14)$\%$, see Table~\ref{tab:112Sn_tau}. The efficiency-calibration problem of Ref.\,\cite{Kum05a} for $\gamma$-decays with $E_{\gamma} > 2.45$\,MeV might become even more obvious for the $2^+$ states at 2966.4(3)\,keV and 3092.4(2)\,keV. Our data support the adopted values, while Ref.\,\cite{Kum05a} provided completely opposite results, i.e. a stronger $\gamma$-decay intensity to the ground state. For $\gamma$-decays with $E_{\gamma} < 2.45$\,MeV, as already stated, our results do in general support the previous findings of Ref.\,\cite{Kum05a}, {\it e.g.}, for the $3^+_2$ state at 3077.8(3)\,keV and the $5^-_1$ state at 3132.5(2)\,keV of $^{112}$Sn.
\\

\subsubsection{$^{112}$Sn}

Many new levels and $\gamma$ transitions in $^{112}$Sn were reported in Ref.\,\cite{Kum05a}. We will shortly comment on those levels where conflicting results were observed. However, we also want to stress explicitly that the majority of new levels observed in Ref.\,\cite{Kum05a} is supported by our data, compare Table~\ref{tab:112Sn_tau}.
\\

{\it 3141\,keV:} A new level was proposed in Ref.\,\cite{Kum05a} based on a $\gamma$ decay with $E_{\gamma} = 990.2(4)$\,keV. This $\gamma$-decay energy does in fact coincide with the one of the $\gamma$-decay of the $4^+_1$ to the $2^+_1$. Based on our data and a careful analysis of different excitation-energy gates we propose to reject this level assignment and claim that this $\gamma$-ray has been solely observed due to feeding in Ref.\,\cite{Kum05a}. Two excited states in the relevant energy range, i.e. at 3113.2(2)\,keV and 3132.5(2)\,keV decay to the $4^+_1$ state.
\\

{\it 3288\,keV:} Also this level was proposed in Ref.\,\cite{Kum05a}. The assignment was based on the observation of a $\gamma$ transition with an energy of 1097.2(3)\,keV which coincides with the $\gamma$-decay energy of the $3^-_1$ state to the $2^+_1$, compare Table~\ref{tab:112Sn_tau}. We propose to reject this assignment due to feeding. Two excited states at 3248.2(2)\,keV and 3285.7(2)\,keV decay to the $3^-_1$ state in the relevant energy interval. Both states were observed in the $(n,n'\gamma)$ experiment of Ref.\,\cite{Kum05a} as well.
\\

{\it 3524\,keV:} While the $\gamma$ transition with an energy of 3524.2(10)\,keV was proposed to belong to a $J^{\pi} = 2^+$ state at 3524.3(3)\,keV in Ref.\,\cite{Kum05a}, we propose that two states exist at 3524.0(4)\,keV and 3526.5(2)\,keV, respectively. This observation is supported by both the different level lifetimes observed and the $\gamma$ transitions depopulating the respective levels, see Table~\ref{tab:112Sn_tau}. The lifetime limit of $\tau > 180$\,fs given in Ref.\,\cite{Kum05a} has, thus, to be attributed to the excited state at 3526.5(2)\,keV. For this state a lifetime of 229(27)\,fs has been determined from our data.
\\

\subsubsection{$^{114}$Sn}

Both tin isotopes were studied before by means of the $(p,t)$ reaction\,\cite{Guaz04a, Guaz12a}. Many excited states which were first observed in the $(p,t)$ experiments of Ref.\,\cite{Guaz04a} are now supported by the observation of $\gamma$ decays from these levels, see Table~\ref{tab:114Sn_tau}. Here, we will only comment on the contradicting spin-parity assignments which were made in Ref.\,\cite{Guaz04a} and are partly adopted\,\cite{ENSDF}.
\\
\\

{\it 2514\,keV:} Two levels have been adopted at an energy of 2514\,keV with $J^{\pi} = 3^-$ and $J^{\pi} = 3^+$, respectively. The $3^+$ assignment is based on the measurement of two multipole-mixing ratios $\delta$\,\cite{ENSDF}. Both $\gamma$-decays from and to the $3^+$ state were also observed in our experiment. We, thus, conclude that this state has been excited in our experiment. No signs of the $3^-$ state reported at 2510\,keV\,\cite{Guaz04a} were seen.
\\

{\it 2576\,keV:} Based on the observation of a clear $L = 2$ transfer, a $J^{\pi} = 2^+$ state was proposed at an energy of 2576\,keV\,\cite{Guaz04a}. The level was not observed in our $p\gamma$ coincidence data. However, there is a weak $\gamma$ transition with $E_{\gamma} \sim 390$\,keV in coincidence with the $\gamma$-decay of the $4^+_1$ level seen in our $\gamma\gamma$-coincidence data. Consequently, we cannot exclude the possibility that this state might exist in $^{114}$Sn. However, since the $(p,n)$ channel is open ($Q(p,n) = -6.8$\,MeV), this state is rather populated in the $\beta^+$-decay of $^{114}$Sb to $^{114}$Sn than directly by the $(p,p')$ reaction.
\\

{\it 3025\,keV:} Two excited states are adopted at an energy of 3025\,keV with a spin-parity assignment of $J^{\pi} = 2,3^+$ and $J^{\pi} = 0^+$, respectively. In addition, another close-lying state at 3028\,keV with $J^{\pi} = 2,3^+$ has been reported\,\cite{ENSDF}. In fact, Ref.\,\cite{Guaz04a} reported the $J^{\pi} = 0^+$ assignment for $E_x = 3028(3)$\,keV. Especially, the new $\gamma$ transition to the $0^+_2$ state for the state at 3024.9(2)\,keV excludes a $J^{\pi} = 0^+$ assignment.
\\

{\it 3397\,keV:} Ref.\,\cite{Guaz04a} assigned $J^{\pi} = 6^+$ to the excited state at 3397(3)\,keV. In addition, a possible $4^-$ state is adopted at an energy of 3396.9(5)\,keV which was actually reported to have a $(3,4)^-$ assignment\,\cite{Schim94a,Wir95a}. For the latter a $\gamma$-decay with $E_{\gamma} = 1122.3$\,keV to the $3^-_1$ state was observed to be of mixed $M1+E2$ character. New $\gamma$-decays of this level to the $2^+_1$, $2^+_2$ and $2^+_3$ state have been observed, which favor a $J^{\pi} = 3^-$ spin-parity assignment, see Table~\ref{tab:114Sn_tau}. A comparison of the theoretical $L = 3$ and the experimentally measured angular distribution shown in Ref.\,\cite{Guaz04a} might also support a $J^{\pi} = 3^-$ assignment which is tentatively given in Table~\ref{tab:114Sn_tau}.
\\

{\it 3452\,keV:} Two spin-parity assignments were previously reported for a possible state at $E_x \approx 3452$\,keV, either $6^+$\,\cite{ENSDF} or $0^+$\,\cite{Guaz04a}. The latter is based on a rather clear $L = 0$ transfer seen in the $(p,t)$ reaction. However, it is the $L = 0$ transfer with the smallest cross section observed in Ref.\,\cite{Guaz04a}. The decay of this state to the ground state of $^{114}$Sn has also been observed in an old $(n,n'\gamma)$ experiment\,\cite{ENSDF}. The $6^+$ assignment is, thus, odd. A $\gamma$ decay with an energy of 3452.1(2)\,keV leading to the ground state was also observed in our experiment. It is not possible to originate from a $J^{\pi} = 0^+$ state. Based on our data we propose a $J^{\pi} = 1^-$ spin-parity assignment. As will be discussed in Sec.\,\ref{sec:qoc}, this state is a suitable candidate for the quadrupole-octupole coupled (QOC) $1^-$ state. We cannot exclude the existence of a doublet at this energy and, thus, the existence of an additional $0^+$ state.
\\ 

{\it 3781\,keV:} This $2^+$ state is strongly populated in the $\beta^+$-decay of $^{114}$Sb to $^{114}$Sn\,\cite{Wig76a}, i.e. in the $(p,n)$ reaction and is, thus, also clearly seen in our $\gamma\gamma$-coincidence data. However, it is not observed at all in our $p\gamma$-coincidence data, i.e. it is not strongly excited in the $(p,p')$ reaction.

\section{discussion}

\subsection{Shape coexistence and multiphonon quadrupole states in $^{112,114}$Sn}

\begin{figure}[t]
\centering
\includegraphics[width=1\linewidth]{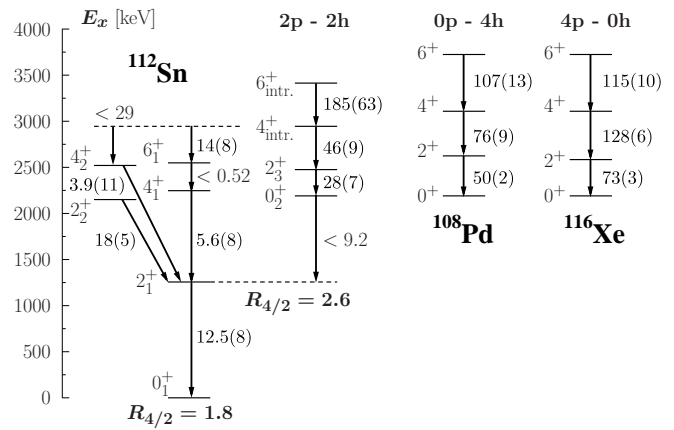}
\caption{\label{fig:sc_112Sn}Normal and 2p-2h intruder states in $^{112}$Sn. The $B(E2)\downarrow$ values are given in W.u.. The data for the intruder $6^+$ state has been taken from Ref.\,\cite{Gangu01a}. The 0p-4h ($^{108}$Pd) and 4p-0h ($^{116}$Xe) Yrast sequences are also shown and have been shifted to the energy of the $0^+_2$ state of $^{112}$Sn. The $B(E2;2^+_1 \rightarrow 0^+_1)$ value is taken from Ref.\,\cite{Jungcl11a}. The data for $^{108}$Pd and $^{116}$Xe was taken from Ref.\,\cite{ENSDF}.}
\end{figure}

\begin{figure}[t]
\centering
\includegraphics[width=1\linewidth]{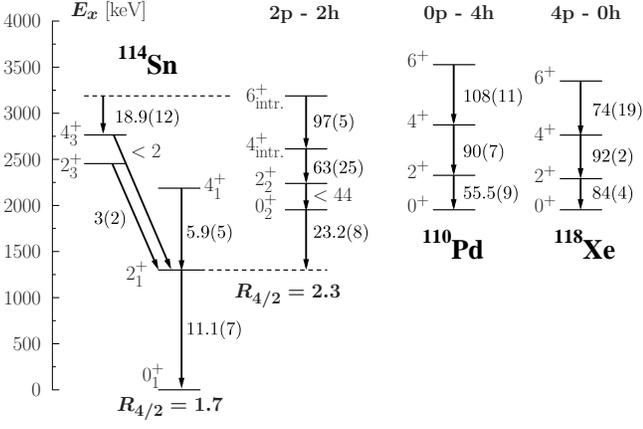}
\caption{\label{fig:sc_114Sn}Same as Fig.\,\ref{fig:sc_112Sn} but for $^{114}$Sn. The $B(E2;2^+_1 \rightarrow 0^+_1)$ value is taken from Ref.\,\cite{Jungcl11a} and the data for the intruder $6^+$ state was taken from Ref.\,\cite{Gabl01a}. The $^{110}$Pd and $^{118}$Xe was taken from Ref.\,\cite{ENSDF}.}
\end{figure}

Shape coexistence has been discussed in the Sn isotopes for decades, see the review articles\,\cite{Wood92a, Heyd11a}, and its existence has been mainly attributed to proton 2p-2h excitations across the $Z = 50$ shell closure. Many experiments have been performed to study the positive-parity and negative-parity intruder bands in $^{112,114}$Sn, see, {\it e.g.}, Refs.\,\cite{Hara88a, Schim92a, Gabl01a, Gangu01a}. We were now able to establish the $0^+$, $2^+$ and $4^+$ members of the positive-parity intruder configuration in $^{112,114}$Sn, see Figs.\,\ref{fig:sc_112Sn} and \ref{fig:sc_114Sn}. These are clearly identified in terms of their interband transitions which are by far the most collective $E2$ transitions. For instance, the $B(E2;2^+_3 \rightarrow 0^+_1)$ and $B(E2;2^+_3 \rightarrow 2^+_1)$ values amount only to 0.082(13)\,W.u. and 0.23(6)\,W.u. in $^{112}$Sn, respectively. The transition to the $2^+_1$ as well as the $B(E2)$ value to the $2^+_2$ are also weak for the intruder $4^+_{\mathrm{intr.}}$ state. However, strong $E2$ transitions are observed to both the $6^+_1$ and $4^+_2$ state indicating that these states might be structurally related, see Fig.\,\ref{fig:sc_112Sn}. It must be mentioned that for the latter transition a multipole-mixing ratio needs to be determined to make a final statement. Interestingly, a very collective $E2$ strength of $B(E2;6^+_{\mathrm{intr.}} \rightarrow 4^+_{2784\,\mathrm{keV}}) = 68(22)$\,W.u. is calculated based on the data of Refs.\,\cite{Gangu01a,ENSDF}. The $B(E2)\downarrow$ to the $4^+_2$ is 9(4)\,W.u. and, thus, comparable to the $B(E2; 4^+_{\mathrm{intr.}} \rightarrow 6^+_1)$ value, see Fig.\,\ref{fig:sc_112Sn}.
\\

In $^{114}$Sn, the $4^+_{\mathrm{intr.}}$ state corresponds to the $4^+_2$ state at 2613.7(4)\,keV, see Fig.\,\ref{fig:sc_114Sn}. Therefore, the decays seen and discussed in $^{112}$Sn are not observed. However, in contrast to $^{112}$Sn, the $B(E2;4^+_{\mathrm{intr.}} \rightarrow 2^+_1) = 6.6(10)$\,W.u. is comparably large. Interestingly, this is also true for the $B(E2;2^+_{\mathrm{intr.}} \rightarrow 2^+_1)$ value which is $\leq 8$\,W.u.. The $6^+_{\mathrm{intr.}}$ state has been identified at an energy of 3188\,keV\,\cite{Gabl01a}, see also Fig.\,\ref{fig:sc_114Sn}. The $B(E2; 6^+_{\mathrm{intr.}} \rightarrow 4^+_3) = 18.9(12)$\,W.u. is also comparably large. Still, it is at least a factor of 2 smaller than the value calculated for the corresponding transition to the $4^+$ state at 2783.5(2)\,keV in $^{112}$Sn.
\\

\begin{table}[t]
\centering
\caption{\label{tab:ibm_114sn}Comparison of the normal and intruder configurations identified experimentally and the predictions of the $sd$ IBM-2 with mixing in $^{114}$Sn. The parameters for the intruder configuration were adopted from Ref.\,\cite{Kim96a}, i.e. $^{110}$Pd. The parameters for the normal configuration in $^{114}$Sn were adopted from Ref.\,\cite{Singh97a} but slightly changed, i.e. $C_{0\nu} = -0.55$, $C_{2\nu} = 0$, and $C_{4\nu} = -0.31$. The mixing parameters $\alpha$ and $\beta$ were kept at 0.2 and 0, respectively. $\Delta$, i.e. the relative energy shift between the normal and intruder configurations was set to 2.78\,MeV. The parameters of the $E2$ operator were also slightly changed to $e_{\nu} = 0.07\,\mathrm{eb^2}$, $e_{\pi} = 0.105\,\mathrm{eb^2}$ and $e_2/e_0 = 1.43$. The experimental $B(E2;2^+_1 \rightarrow 0^+_1)$ value is taken from Ref.\,\cite{Jungcl11a}. For a description of the Hamiltonian, the $E2$ operator and their parameters see, {\it e.g.}, Refs.\,\cite{Del93a, Kim96a}.}
\begin{ruledtabular}
\begin{tabular}{cccccc}
$J^{\pi}_i$ & $E_x$ & $E_{x,\mathrm{IBM}}$ & $J^{\pi}_f$ & $B(E2)_{\mathrm{exp.}}\downarrow$ & $B(E2)_{\mathrm{IBM}}\downarrow$ \\
          & [MeV] & [MeV]                &       & [W.u.]                  & [W.u.] \\
\hline
\multicolumn{6}{c}{normal configuration} \\
\hline
$2^+_1$ & 1.30 & 1.30 & $0^+_1$ & 11.1(7) & 11 \\
$4^+_1$ & 2.19 & 2.28 & $2^+_1$ & 5.9(5) & 19 \\
$0^+_2$ & 1.95 & 1.99 & $2^+_1$ & 23.2(8) & 21 \\
$2^+_3$ & 2.45 & 2.54 & $0^+_1$ & 0.023(9) & 0.004 \\
        &      &      & $2^+_1$ & 3(2) & 17 \\
        &      &      & $2^+_2$ & -    & 8 \\
\hline
\multicolumn{6}{c}{intruder configuration} \\
\hline
$0^+_3$ & 2.16 & 2.15 & $2^+_1$ & $\leq 5$ & 2 \\
$2^+_2$ & 2.24 & 2.46 & $0^+_1$ & $\leq 0.12$ & 0.04 \\
        &      &      & $2^+_1$ & $\leq 8$ & 2 \\
        &      &      & $0^+_2$ & $\leq 44$ & 31 \\
        &      &      & $0^+_3$ & -         & 27 \\
$4^+_2$ & 2.61 & 3.00 & $2^+_1$ & 6.6(10) & 0.2 \\
        &      &      & $4^+_1$ & 1.6(10) & 0.06 \\
        &      &      & $2^+_2$ & 62(25)  & 85 \\
$6^+$   & 3.19 & 3.63 & $4^+_1$ & 1.68(9) & 1.5\\
        &      &      & $4^+_2$ & 97(5)   & 93\\
        &      &      & $4^+_3$ & 18.9(12)& 0.7\\
\end{tabular}
\end{ruledtabular}
\end{table}

To test the mixing hypothesis between the normal and intruder configuration, we performed $sd$ IBM-2 calculations using the computer code NPBOS\,\cite{Otsu85a} for $^{114}$Sn, see Table\,\ref{tab:ibm_114sn}. As can be seen in Figs.\,\ref{fig:sc_112Sn} and \ref{fig:sc_114Sn}, the observed reduced $E2$ transition strengths are closer to the corresponding quantities observed in the Pd isotopes, i.e. the 0p-4h nucleus. Similar observations were made for the $B(E2;2^+_{\mathrm{intr.}} \rightarrow 0^+_{\mathrm{intr.}})$ values in the Cd isotopes with $N = 62 - 68$ which are comparable to the $B(E2;2^+_1 \rightarrow 0^+_1)$ values observed in the corresponding Ru isotopes\,\cite{Hey92a}. We, therefore, adopted the IBM-2 parameters of Ref.\,\cite{Kim96a} determined for $^{110}$Pd to describe the intruder configuration. For the normal configuration we slightly adjusted the parameters which were reported in Ref.\,\cite{Singh97a}, see the caption of Table~\ref{tab:ibm_114sn}. As can be seen the $B(E2)$ strengths in the intruder band are nicely described by the model. Also the ``interband'' transitions are fairly well described. Of course, deviations are observed, see, {\it e.g.}, the $B(E2;4^+_1 \rightarrow 2^+_1)$ and $B(E2;2^+_3 \rightarrow 2^+_1)$ values. It is tempting to speculate that these deviations arise from mixing effects which are not covered by the simplified IBM approach or if the mixing parameter $\beta$ is kept at 0. Clearly, certain configurations will be outside of the $sd$ IBM-2 model space. In fact, indications of mixing effects between the two $4^+$ states at 2613.7(4)\,keV and 2764.9(5)\,keV were already proposed based on their excitation energies, see the review article\,\cite{Wood92a}. These are now further strengthened by reduced $E2$ transition strengths. Still, the decay rates of the $0^+_2$ and $0^+_3$ as well as the decay rate to the $0^+_2$ are perfectly described using the IBM approach. These two $0^+$ states are almost perfectly mixed. The $0^+_3$ has a slightly larger admixture of the intruder configuration. It is, thus, not surprising that the two $B(E2; 2^+_{\mathrm{intr.}} \rightarrow 0^+_{i})$ values ($i =2,3$) add up to the $B(E2;2^+_1 \rightarrow 0^+_1)$ value observed in $^{110}$Pd, compare Fig.\,\ref{fig:sc_114Sn}. Unfortunately, the decay $2^+_2 \rightarrow 0^+_3$ has not been observed in $^{114}$Sn so far. If the scenario drawn is true a $\gamma$-decay intensity $I_{\gamma}$ of about 0.002\,$\%$ would be expected. Indeed, the $2^+_2 \rightarrow 0^+_3$ was recently observed for the case of $^{116}$Sn\,\cite{Pore17a}, which was populated through the $\beta^-$-decay of $^{116m1}$In. Here, an $I_{\gamma}$ of 0.0091(6)\,$\%$ was determined and a very collective $B(E2)\downarrow$ of 100(8)\,W.u. was calculated. Given the adopted value is correct, the $B(E2;2^+_1 \rightarrow 0^+_1)$ is 41(6)\,W.u. in $^{112}$Pd. A summed $B(E2; 2^+_{\mathrm{intr.}} \rightarrow 0^+_i)$ strength of 144(8)\,W.u. would consequently not be expected if the intruder configuration would solely result from $^{112}$Pd in $^{116}$Sn. Unfortunately, lifetimes of the $4^+_{\mathrm{intr.}}$ and $6^+_{\mathrm{intr.}}$ are not known. A stringent comparison is presently not possible. It is still interesting to note that in contrast to $^{112}$Sn the $B(E2;0^+_2 \rightarrow 2^+_1) = 18(2)$\,W.u. is similar to the one observed in $^{114}$Sn. For a deeper understanding of mixing between possible two-phonon quadrupole states and intruder states further investigations are clearly necessary. 
\\

\begin{table}[t]
\centering
\caption{\label{tab:multph_sn}Possible candidates for three-phonon quadrupole states in $^{112,114}$Sn. If limits for the $B(E2)$ values are given, either the multipole-mixing ratio $\delta$, the specific lifetime $\tau$ or both quantities are unknown, see Tables~\ref{tab:112Sn_tau} and \ref{tab:114Sn_tau}.}
\begin{ruledtabular}
\begin{tabular}{cccc}
$J^{\pi}_i$ & $E_x$ & $J_f^{\pi}$ & $B(E2)_{\mathrm{exp.}}$ \\
          & [keV] &       & [W.u.]\\
\hline
\multicolumn{4}{c}{$^{112}$Sn} \\
\hline
$0^+$ & 2617.4(3) & $2^+_1$ & $\leq 2$ \\
      &           & $2^+_2$ & $\leq 7$ \\
$2^+$ & 2720.6(2) & $0^+_1$ & $\leq 0.02$ \\
      &           & $2^+_1$ & $0.06^{+0.08}_{-0.01}$ \\
      &           & $2^+_2$ & $\leq 4.3$ \\
      &           & $0^+_2$ & 3.3(12) \\
$3^+$ & 2755.2(3) & $2^+_1$ & $\leq 0.004$\\
      &           & $2^+_2$ & $\leq 12$ \\
      &           & $4^+_1$ & $\leq 45$ \\
      &           & $4^+_2$ & $\leq 0.2$ \\
$4^+$ & 2783.5(2) & $2^+_1$ & 5.1(6) \\
      &           & $4^+_1$ & $\leq 35$ \\
\hline
\multicolumn{4}{c}{$^{114}$Sn} \\
\hline
$4^+$ & 2859.2(5) & $2^+_1$ & 2.8(4) \\
      &           & $4^+_1$ & $\leq 10$ \\
      &           & $2^+_2$ & $< 5$ \\
      &           & $2^+_3$ & $< 46$ \\
$2^+$ & 2943.4(2)  & $0^+_1$ & $< 0.001$ \\
      &           & $2^+_1$ & $\leq 0.3$ \\
      &           & $0^+_2$ & $\leq 0.4$ \\
      &           & $2^+_2$ & $\leq 0.9$ \\
      &           & $0^+_4$ & $\leq 1.6$ \\
      &           & $2^+_3$ & $\leq 5.2$ \\
$0^+$ & 3028.0(2) & $2^+_1$ & 1.7(7) \\
      &           & $2^+_2$ & 1.4(10) \\
      &           & $2^+_3$ & 16(8) \\
\end{tabular}
\end{ruledtabular}
\end{table}

\begin{figure}[t]
\centering
\includegraphics[width=0.98\linewidth]{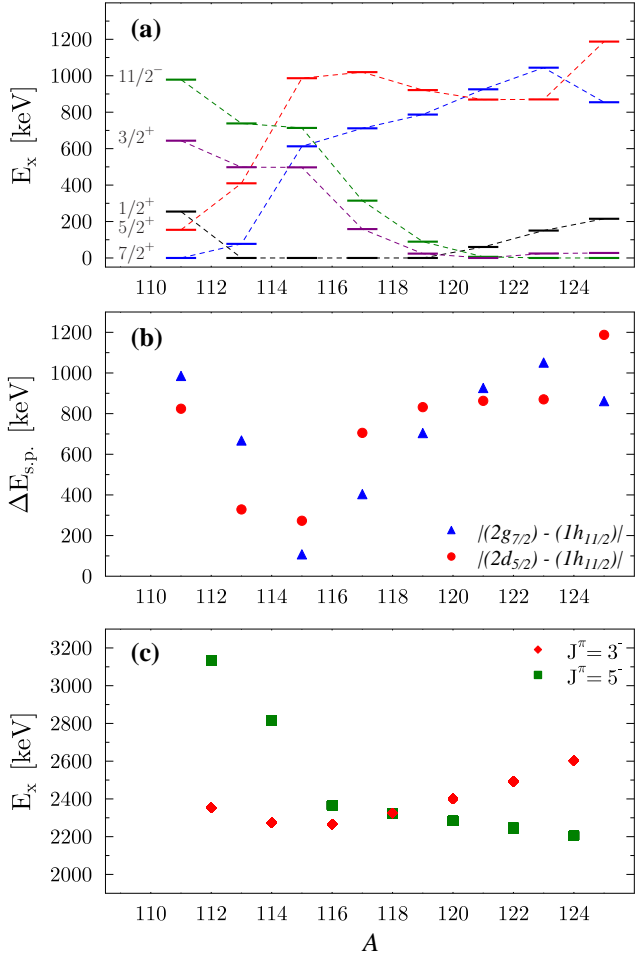}
\caption{\label{fig:odd_Sn}(color online) {\bf (a)} systematics of the low-lying single-particle states in the odd Sn isotopes. The data have been compiled from Ref.\,\cite{ENSDF}. The experimentally observed states might in good approximation reflect the corresponding single-particle levels, i.e. $1g_{7/2}$ (blue), $2d_{5/2}$ (red), $3s_{1/2}$ (black), $2d_{3/2}$ (purple), and $1h_{11/2}$ (green) . {\bf (b)} energy difference between the $1h_{11/2}$ level and the $2d_{5/2}$ as well as the $2g_{7/2}$. {\bf (c)} energy evolution of the $3^-_1$ (red diamonds) and $5^-_1$ state (green squares) in the stable even-even Sn isotopes\,\cite{ENSDF}. For the discussion, see text.}
\end{figure}

As stressed in the introduction, candidates for three-phonon quadrupole states were identified in $^{124}$Sn\,\cite{Band05a}. Possible candidates in $^{112,114}$Sn are given in Table~\ref{tab:multph_sn}. The experimentally calculated $B(E2)$ values are indeed similar to the values which were observed in $^{124}$Sn, i.e. the forbidden transitions are approximately weaker by one order of magnitude compared to the transitions leading to the two-phonon states. Note, that for most states only upper limits could be determined. We also have to keep in mind that the two-phonon states are not pure, compare Table~\ref{tab:ibm_114sn}. In addition, at least one other configuration is present in $^{114}$Sn, i.e. $(3s_{1/2})^{-1}(1g_{7/2})^{1}$ leading to $J^{\pi} = 3^+$ and $4^+$. The $B(M1;4^+ \rightarrow 3^+_1) \approx 0.1$\,$\mu_N^2$ between the states at 2764.9(5)\,keV and 2514.4(2)\,keV is the largest value observed and might be caused by the corresponding spin-flip transition. Whether the proposed $3^+$ and $4^+$ three-phonon quadrupole members have the same admixture is not clear. Clearly, the structure of the ground state, i.e. the underlying single-particle structure changes from $^{112}$Sn to $^{114}$Sn. The main fragments of the single-particle levels in the odd-A Sn isotopes can be seen in Fig.\,\ref{fig:odd_Sn}\,{\bf (a)}. It will be discussed in connection with the negative-parity states. For unambiguous assigments, multipole-mixing ratios $\delta$ need to be determined in the future. For now, we can only conclude that the states discussed do not decay as expected for pure three-phonon states and that the situation in $^{114}$Sn seems to be even more complex. The latter might be attributed to the small $N = 64$ subshell gap which is also seen in Fig.\,\ref{fig:odd_Sn}\,{\bf (a)} and even more pronounced mixing effects.

\subsection{Quadrupole-octupole coupled states}
\label{sec:qoc}

\subsubsection{The $J^{\pi} = 1^-$ candidate}

\begin{figure}[!t]
\centering
\includegraphics[width=0.75\linewidth]{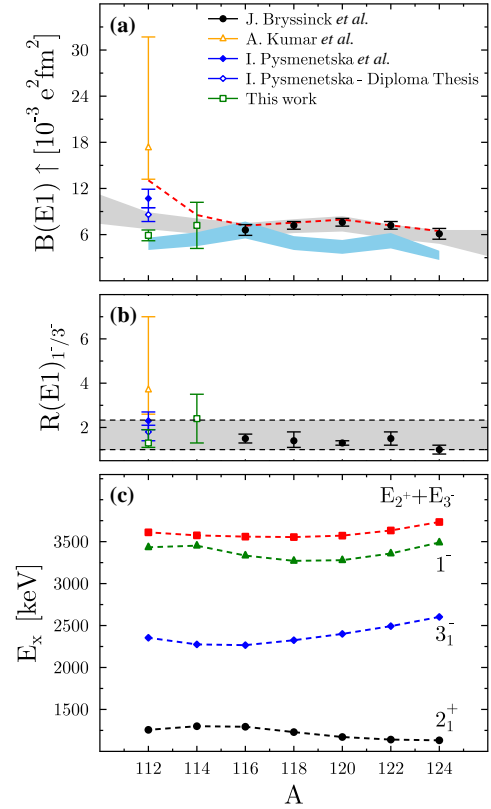}
\caption{\label{fig:e1_Sn}(color online) Systematics for the two-phonon $1^-$ state in the stable Sn isotopes. {\bf (a)} $B(E1)$\,$\uparrow$ values determined in Refs.\,\cite{Bryss99a,Pys04a,Pys06a,Kum05a} and this work. The grey band corresponds to the $B(E2; 0^+_1 \rightarrow 2^+_1)$ values of Ref.\,\cite{Jungcl11a} scaled to the $E1$ strength of $^{116}$Sn, i.e. assuming $B(E1) \sim \beta_2$, while the light-blue band corresponds to the $B(E3; 0^+_1 \rightarrow 3^-_1)$ values of Ref.\,\cite{Jon81a} scaled in the same way, i.e. $B(E1) \sim \beta_3$. The RQTBA calculations are shown as well (red dashed line)\,\cite{Lit10a, Lit13b}. Note that these results were scaled with a factor of 0.5, see text. {\bf (b)} R(E1)$_{1^-/3^-}$\,=\,$B(E1;1^- \rightarrow 0^+_1)$/$B(E1;3^-_1 \rightarrow 2^+_1)$ and a band of expected values for a two-phonon structure in grey, see Ref.\,\cite{Piet99a}. {\bf (c)} Experimentally determined excitation energies of the two-phonon $1^-$ candidates and the excitation energies of the constituent-phonon states, respectively.}
\end{figure}

The quadrupole-octupole coupled $1^-$ state has been systematically studied in $^{116-124}$Sn using the NRF technique\,\cite{Bryss99a}. Later on, $^{112}$Sn was added to the systematics\,\cite{Pys04a,Pys06a} including the aforementioned $(n,n'\gamma)$ experiment\,\cite{Kum05a}. The last missing stable Sn isotope, i.e. $^{114}$Sn, was added in this work. Figs.\,\ref{fig:e1_Sn}\,{\bf (a)}-{\bf (c)} present the existing and new data. The two-phonon $1^-$ candidate in $^{114}$Sn at 3452\,keV fits well into the Sn systematics in terms of the {\bf (a)} $B(E1;0^+_1 \rightarrow 1^-)$ strength, {\bf (b)} $R(E1)$ ratio and {\bf (c)} energy systematics. As seen in all other Sn isotopes, its energy lies slightly below the sum energy of the constituent phonons and seems to be more sensitive to the evolution of the excitation energy of the $2^+_1$ state, i.e. a shallow maximum is observed, compare Fig.\,\ref{fig:e1_Sn}\,{\bf (c)}. Note that no other suitable candidate is observed in the relevant energy range. Furthermore, as in all other Sn isotopes, no $\gamma$ decay besides the ground-state decay is observed. It shall be mentioned that the comparably large uncertainty for the lifetime of the $^{114}$Sn candidate is caused by the large $F(\tau)$ value of 0.93(3) and its proximity to unity. The situation in $^{112}$Sn remains unsatisfying. Although the candidate is clearly identified, none of the measurements are found in agreement with any other measurement in terms of the $E1$ strength, see Fig.\,\ref{fig:e1_Sn}\,{\bf (a)}. The value determined by A. Kumar {\it et al.}, however, seems to be too large. It does not match the empirically determined range of two-phonon $B(E1)$ strengths and its proposed connection to the $B(E1;3^-_1 \rightarrow 2^+_1)$ value\,\cite{Piet99a}, see Fig.\,\ref{fig:e1_Sn}\,{\bf (b)}. Since there were also ambiguities in the efficiency determination for the $^{112}$Sn$(\gamma,\gamma')$ experiment\,\cite{Pys04a}, no decision in favor of any of the remaining experiments can be made. The $E1$ strength, see Fig.\,\ref{fig:e1_Sn}\,{\bf (a)}, seems to follow the evolution of the $B(E2;0^+_1 \rightarrow 2^+_1)$ value (grey-shaded area) rather than the evolution of the $B(E3;0^+_1 \rightarrow 3^-_1)$ value (light-blue shaded area), which might hint at a common origin of the strength. This should be investigated further using a stringent comparison to theory. Calculations in the framework of the RQTBA have already been performed for $^{112,116,120,124}$Sn\,\cite{Lit10a, Lit13b} and new calculations for $^{114,118,122}$Sn were added in this work. Unfortunately, the RQTBA overestimates the experimental values by a factor of about 2. In Fig.\,\ref{fig:e1_Sn}\,{\bf (a)}, the theoretical results have been scaled with this factor and are shown as a red dashed line. Note, that, even though a two-phonon structure is predicted by the RQTBA, the origin of the strength evolution and especially the strength increase for $^{112}$Sn are presently not understood. One should mention that the RQTBA is a QCD-based self-consistent approach, which does not involve any adjustment of parameters besides the meson masses and meson couplings. These are fitted to global nuclear properties. In this theory, the two-phonon states are considered as tiny structures and an agreement within a factor of 2 is often considered sufficient. However, future advancements of the RQTBA, i.e. the implementation of higher-order correlations are expected to improve the agreement with experiment. Presently, the RQTBA is limited to phonon+2QP configurations. The excitation energy of the $1^-$ two-phonon state is also rather approximate in the RQTBA since it appears to be quite sensitive to the pairing strength. Still, in fair agreement with experiment the QOC $1^-$ is predicted between 2.66\,MeV ($^{116}$Sn) and 3.98\,MeV ($^{124}$Sn) in the stable Sn isotopes. The candidates in $^{112}$Sn and $^{114}$Sn are predicted at 3.85\,MeV and 3.70\,MeV, respectively.
\\

Interestingly, additional candidates for the $1^-$ QOC state might be observed around the expected energy, see Tables~\ref{tab:qoc_112sn} and \ref{tab:qoc_114sn}. Unfortunately, a conclusive spin-parity assignment is not possible at the moment. A $J^{\pi} = 2^+$ assignment would be possible as well. Still, the additional candidates in $^{112}$Sn and $^{114}$Sn show similar decay properties, i.e. small $B(E1)$ values and $B(E2;(1^-) \rightarrow 3^-_1) \approx 6$\,W.u.. As will be discussed in the next part, such $B(E2)$ values are in fact expected for the members of the QOC quintuplet. 

\subsubsection{The other quintuplet members}

\begin{table*}[!t]
\caption{\label{tab:qoc_112sn}Candidates for the quadrupole-octupole coupled states in $^{112}$Sn. The excitation energy is shown in the first column, while the spin-parity assignment is given in the second column. The third and fourth column specify the final state, where the decay with a specific $\gamma$ energy and branching ratio, see fifth and sixth column, is leading to. The last two columns present the $E1$ and $E2$ reduced transition probabilities, respectively. The $B(E3;3^-_1 \rightarrow 0^+_1)$ corresponds to 17(2)\,W.u. in $^{112}$Sn\,\cite{Jon81a}, respectively. No multipole mixing ratios $\delta$ could be determined in this work. Pure transitions have been assumed where no $\delta$ was previously known.}
\begin{ruledtabular}
\begin{tabular}{ccccccccc}
$E_x$& $J^{\pi}$ & $J^{\pi}_f$ & $E_f$ & $E_{\gamma}$ & $I_{\gamma}$ & B(E1)$\downarrow$ & B(E2)$\downarrow$ \\
$\mathrm{[keV]}$ & & & [keV] & [keV] & & [mW.u.] & [W.u.]\\
\hline
1256.5(2) & $2^+_1$ & $0^+_1$ &0 & 1256.5(2) & 1 & - & 12.5(7)\footnote{Ref.\,\cite{Jungcl11a}} \\
2353.7(2) & $3^-_1$ & $2^+_1$ & 1256.5(2) & 1097.2(2) & 1 & 1.13(8) & - \\
\hline
\multirow{3}{*}{3383.3(2)} & \multirow{3}{*}{$3^-$} & $2^+_1$ &1256.5(2) & 2126.8(2) & 0.85(2) & 0.120(9) & - \\
 &  &  $2^+_2$ & 2150.5(3) & 1232.9(2) & 0.041(9) & 0.030(7) & - \\
  &  & $(2^+,3,4^+)$ & 2917.0(2) & 466.5(2) & 0.11(2) & 1.5(2) & - \\
  \hline
\multirow{4}{*}{3396.6(2)} & \multirow{4}{*}{$2^{(-)}$} & $2^+_1$ &1256.5(2) & 2139.9(2) & 0.057(13) & 0.005(2) & - \\
 &  &  $2^+_2$ & 2150.5(2) & 1246.1(2) & 0.64(3) & 0.30(9) & - \\
  &  & $3^-_1$ & 2353.7(2) & 1042.4(2) & 0.27(5) & - & 9$^{+3}_{-7}$ \\
    &  & $2^+_4$ & 2720.6(2) & 675.8(2) & 0.039(9) & 0.11(5) & - \\
    \hline
3433.4(2) & $1^{(-)}$ & $0^+_1$ & 0 & 3433.4(2) & 1 & 1.31(15) & - \\
\hline
\multirow{3}{*}{3497.9(2)} & \multirow{3}{*}{$5^-$} & $3^-_1$ &2353.7(2) & 1144.2(2) & 0.70(4) & - & 29(13) \\
 &  &  $4^+_2$ & 2520.5(2) & 977.1(2) & 0.27(6) & 0.39(18) & - \\
  &  & $4^+$ & 2783.5(2) & 714.7(3) & $\leq 0.03$ & $\leq 0.19$ & - \\
  \hline
  \multirow{2}{*}{3553.2(2)} & \multirow{2}{*}{$(3)^-$} & $2^+_1$ &1256.5(2) & 2296.8(2) & 0.83(3) & 0.06(2) & - \\
 &  &  $3^+_1$ & 2755.2(3) & 797.7(3) & 0.17(3) & 0.30(10) & - \\
 \hline
\multirow{3}{*}{3827.1(3)} & \multirow{3}{*}{$(1^-,2^+)$} & $0^+_1$ &0 & 3827.1(2) & 0.58(3) & 0.040(5) & - \\
 &  &  $2^+_1$ & 1256.5(2) & 2570.8(2) & 0.29(5) & 0.066(11) & - \\
  &  & $3^-_1$ & 2353.7(2) & 1473.0(7) & 0.13(3) & - & 4.5(11) \\
\hline
\multirow{3}{*}{3984.7(3)} & \multirow{3}{*}{$(1^-,2^+)$} & $0^+_1$ &0 & 3984.7(3) & 0.79(2) & 0.08(2) & - \\
 &  &  $3^-_1$ & 2353.7(2) & 1630.0(3) & 0.14(2) & - & 5(2) \\
  &  & $2^+_3$ & 2475.5(2) & 1507.8(4) & 0.07(2) & 0.137(95) & - \\
\end{tabular}
\end{ruledtabular}
\end{table*}

\begin{table*}[!t]
\caption{\label{tab:qoc_114sn}Same as Table~\ref{tab:qoc_112sn} but for $^{114}$Sn. The $B(E3;3^-_1 \rightarrow 0^+_1)$ corresponds to 19(2)\,W.u. in $^{114}$Sn\,\cite{Jon81a}.}
\begin{ruledtabular}
\begin{tabular}{ccccccccc}
$E_x$& $J^{\pi}$ & $J^{\pi}_f$ & $E_f$ & $E_{\gamma}$ & $I_{\gamma}$ & B(E1)$\downarrow$ & B(E2)$\downarrow$ \\
$\mathrm{[keV]}$ & & & [keV] & [keV] & & [mW.u.] & [W.u.]\\
\hline
1299.7(2) & $2^+_1$ & $0^+_1$ & 0 & 1299.7(2) & 1 & - & 11.1(7)\footnote{Ref.\,\cite{Jungcl11a}} \\
2274.5(2) & $3^-_1$ & $2^+_1$ & 1299.7(2) & 974.8(2) & 1 & 0.65(8) & - \\
\hline
\multirow{2}{*}{2814.6(2)} & \multirow{2}{*}{$5^-_1$} & $4^+_1$ & 2187.3(3) & 627.4(2) & 0.88(2) & $\leq 0.77$ & - \\
 &  &  $3^-_1$ & 2274.5(2) & 539.9(2) & 0.12(3) & - & $\leq 38$ \\
 \hline
\multirow{4}{*}{2904.9(3)} & \multirow{4}{*}{$3^-$} & $2^+_1$ & 1299.7(2) & 1605.1(4) & 0.026(5) & 0.0030(14) & - \\
 &  &  $4^+_1$ & 2187.3(3) & 717.3(2) & 0.77(2) & 0.7(3) & - \\
  &  & $3^+_1$ & 2514.4(2) & 390.2(2) & 0.20(3) & 1.6(7) & - \\
    &  & $4^+_2$ & 2613.7(4) & 290.3(4) & 0.011(4) & 0.21(12) & - \\
\hline
\multirow{3}{*}{3225.1(2)} & \multirow{3}{*}{$3^-$} & $2^+_1$ & 1299.7(2) & 1925.4(2) & 0.920(14) & 0.11(2) & - \\
  &  & $2^+_3$ & 2453.8(2) & 771.4(4) & 0.019(7) & 0.04(2) & - \\
 &  &  $3^-$ & 2904.9(3) & 319.9(4) & 0.061(13) & - & - \\
\hline
\multirow{4}{*}{3397.3(2)} & \multirow{4}{*}{$3^-$} & $2^+_1$ & 1299.7(2) & 2097.6(2) & 0.31(5) & 0.06(2) & - \\
 &  &  $2^+_2$ & 2238.6(2) & 1158.3(2) & 0.13(2) & 0.14(6) & - \\
  &  & $3^-_1$ & 2274.5(2) & 1122.0(4) & 0.44(2) & - & $3^{+11}_{-3}$ \\
    &  & $2^+_3$ & 2453.8(2) & 943.2(2) & 0.12(2) & 0.24(10) & - \\
\hline
 3452.1(2) & $(1^{-})$ & $0^+_1$ & 0 & 3452.1(2) & 1 & 1.6(7) & - \\
 \hline
\multirow{3}{*}{3483.9(4)} & \multirow{3}{*}{$(1^-,2^+)$} & $2^+_1$ & 1299.7(2) & 2184.1(2) & 0.671(13) & 0.06(2) & - \\
  &  & $0^+_3$ & 2155.9(2) & 1327.7(3)\footnote{No clear assignment possible, see Table\,\ref{tab:114Sn_tau}.} & 0.094(14) & 0.037(11) & - \\
 &  &  $3^-_1$ & 2274.5(2) & 1209.0(2) & 0.235(14) & - & 5.2(13) \\
\hline 
 \multirow{3}{*}{3514.1(3)} & \multirow{3}{*}{$3^-$} & $2^+_1$ & 1299.7(2) & 2214.4(2) & 0.76(3) & 0.14(6) & - \\
 &  &  $4^+_1$ & 2187.4(3) & 1327.0(4)\footnote{No clear assignment possible, see Table\,\ref{tab:114Sn_tau}.} & 0.07(2) & 0.06(3) & - \\
  &  & $2^+_2$ & 2238.6(2) & 1275.0(3) & 0.17(3) & 0.17(8) & - \\
\hline
\multirow{4}{*}{3524.4(2)} & \multirow{4}{*}{$3^-$} & $2^+_1$ & 1299.7(2) & 2224.5(3) & 0.55(3) & 0.023(17) & - \\
 &  &  $4^+_1$ & 2187.3(3) & 1158.3(2) & 0.15(3) & 0.028(22) & - \\
  &  & $3^-_1$ & 2274.5(2) & 1122.0(4) & 0.13(2) & - & 1.2(10) \\
    &  & $2^+_3$ & 2453.8(2) & 943.2(2) & 0.17(3) & 0.08(6) & - \\
\hline
3610.2(4) & $5^{(-)}$ & $4^+_1$ & 2187.3(3) & 1422.9(3) & 1 & 1.0(3) & - \\
\hline
\multirow{4}{*}{3650.3(3)} & \multirow{4}{*}{$(1^-,2^+)$} & $0^+_1$ & 0 & 3650.1(3) & 0.44(3) & 0.012(4) & - \\
 &  &  $2^+_1$ & 1299.7(2) & 2350.3(3) & 0.11(2) & 0.011(5) & - \\
  &  & $0^+_3$ & 2155.9(2) & 1493.7(3) & 0.09(2) & 0.04(2) & - \\
    &  & $3^-_1$ & 2274.5(2) & 1374.6(2) & 0.36(6) & - & 6(2) \\
\end{tabular}
\end{ruledtabular}
\end{table*}

If the two-phonon interpretation is correct, a quintuplet of negative-parity states, i.e. $(2^+ \otimes 3^-)_{1^--5^-}$, should be observed close to the sum energy of the constituent-phonon states. In $^{112}$Sn this sum energy is 3.61\,MeV and in $^{114}$Sn it is 3.57\,MeV. Furthermore, these coupled states should approximately decay according to the properties of their constituent phonons: 

\begin{align*}
B(E2;(2^+_1 \otimes 3^-_1) \rightarrow 3^-_1) = B(E2;2^+_1 \rightarrow 0^+_1) \\
B(E3;(2^+_1 \otimes 3^-_1) \rightarrow 2^+_1) = B(E3;3^-_1 \rightarrow 0^+_1)
\end{align*}

For the case of $^{112}$Sn, candidates have already been proposed in Ref.\,\cite{Kum05a}. The possible candidates for $^{112,114}$Sn, which have been observed in this work, are shown in Tables~\ref{tab:qoc_112sn} and \ref{tab:qoc_114sn}. Despite the $5^-$ state at 3.1\,MeV, which was discussed as a member of the multiplet in Ref.\,\cite{Kum05a}, the $2^-$ and $3^-$ candidates at about 3.4\,MeV were also observed in our experiment. For the case of the tentatively assigned $2^-$ state at 3396.6(2)\,keV, the decay to the $3^-_1$ state was observed and the $B(E2)\downarrow$ agrees with the expectations. The rather small $E1$ transition rate to the $2^+_1$ state might hint at a non-negligible $E3$ or $M2$ contribution. However, assuming a pure $E3$ character of this transition results in an unphysically large value. Certainly, multipole mixing ratios should be determined. 
\\

In general, the unnatural-parity states are only weakly excited in the present experiments, i.e. besides the $1^-$ multiplet candidates, mainly candidates for the $3^-$ and $5^-$ state have been observed. The observed $E2$ decay rate from the $5^-$ state at 3497.9(2)\,keV to the $3^-_1$ state also matches the expectations, while no such $\gamma$-decay branching could be observed for any $3^-$ candidate in $^{112}$Sn. The situation in $^{114}$Sn is reversed. Two $3^-$ states are observed close to the sum energy which decay to the $3^-_1$ state. However, only the $B(E2)$ value of the state at 3397.3(2)\,keV might allow a QOC interpretation within its comparably large uncertainties. Interestingly, the $B(E1;3^- \rightarrow 2^+_1)$ value is one order of magnitude smaller than the $B(E1;3^-_1 \rightarrow 2^+_1)$ value. This might raise the question whether enhanced $E1$ transitions are indeed expected for QOC candidates or if other mechanisms and structures have to be considered. The $B(E1;5^- \rightarrow 4^+_1)$ of the state at 3610.2(4)\,keV is as large as the $B(E1;3^-_1 \rightarrow 2^+_1)$ value, see Table~\ref{tab:qoc_114sn}. 
\\

In this context, it is necessary to mention that for the case of $^{112}$Cd the identification of the $5^-$ multiplet candidates in terms of $E2$ transition rates to the $3^-_1$ state as well as in terms of energy arguments\,\cite{Garr99a} was certainly not sufficient. These states exhibited strong neutron single-particle character as was shown in a $(d,p)$ reaction leading to excited states of $^{112}$Cd\,\cite{Jami14a}. The $5^-$ state at 2373\,keV was interpreted to have a dominant $3s_{1/2} \otimes 1h_{11/2}$ configuration and to be a rotational-band member of the $3^-_1$ one-octupole phonon state which might its explain its enhanced $B(E2;5^- \rightarrow 3^-_1)$ value.
\\

To shed some light, we compiled the excitation energies of the lowest excited states in the odd-$A$ Sn isotopes, see Fig.\,\ref{fig:odd_Sn}\,{\bf (a)}. These states might in good approximation be identified as being the major fragments of the corresponding single-particle levels. In addition, we calculated the energy difference $\Delta E_{s.p.}$ between the $1h_{11/2}$ state and the $2d_{5/2}$ as well as the $1g_{7/2}$ state in Fig.\,\ref{fig:odd_Sn}\,{\bf (b)}, respectively. As can be seen, the energy of $3^-_1$ closely follows both energy differences while the $5^-_1$ state's excitation energy evolves according to the energy of the $1h_{11/2}$ orbital, compare Fig.\,\ref{fig:odd_Sn}\,{\bf (c)} to the other two panels. Since the $3^-_1$ lies below the $5^-_1$ and the necessary quantities are partly known in $^{112-116}$Sn, the $B(E2;5^-_1 \rightarrow 3^-_1)$ can be at least estimated:

\begin{align*}
&^{112}\mathrm{Sn}: B(E2;5^-_1 \rightarrow 3^-_1) \leq 8.2\,\mathrm{W.u.}\\
&^{114}\mathrm{Sn}: B(E2;5^-_1 \rightarrow 3^-_1) \leq 38\,\mathrm{W.u.}\\
&^{116}\mathrm{Sn}: B(E2;5^-_1 \rightarrow 3^-_1) = 2.45(12)\,\mathrm{W.u.}
\end{align*}

We see that assumming pure configurations of $(2d_{5/2})^{-1}(1h_{11/2})^1$ ($\Delta j = \Delta l = 3$) for the $3^-_1$ and $(3s_{1/2})^{-1}(1h_{11/2})^1$ ($\Delta j = \Delta l = 5$) for the $5^-_1$ state might also generate some $E2$ collectivity between the two levels, i.e. the transfer of a valence neutron from the $3s_{1/2}$ orbital to the $2d_{5/2}$ orbital ($\Delta j = \Delta l = 2$) or vice versa. The fact that the $5^-_1$ excitation energy saturates at approximately 2.2\,MeV in the more neutron-rich Sn isotopes where the ground state has a $(1h_{11/2})_{0^+}$ configuration further supports this hypothesis. In a nutshell, approximately 2\,MeV are needed to break a pair and the energy difference between the $1h_{11/2}$ and $3s_{1/2}$ orbital is about 200\,keV in the more neutron-rich stable Sn isotopes, see Fig.\,\ref{fig:odd_Sn}\,{\bf (a)}. However, we want to stress that two things were shown in old shell-model calculations employing a finite-range force and the generalized-seniority scheme with $v \leq 4$, i.e. two broken pairs\,\cite{Bon85a}. First of all, configurations with $v > 2$ and a finite-range force are needed to reproduce the excitation energy and lifetime of the $5^-_1$ state as was shown for $^{112,116}$Sn. These admixtures are on the order of 20\,$\%$. Second, to account for the experimentally observed excitation energy of the $3^-_1$, 1p-1h configurations are needed which require excitations through the $^{100}$Sn inert core. These admixtures could be as large as 43\,$\%$ highlighting the collective nature of the $3^-_1$ state. More modern shell-model calculations which were, however, limited to a $^{100}$Sn inert core support these previous findings\,\cite{Guaz04a, Guaz12a}. 
\\

$^{112}$Cd and $^{114}$Sn have both $N = 64$ and similar neutron components are expected to contribute. The $5^-_1$ state critically discussed in $^{112}$Cd might, thus, have the same structure as the $5^-_1$ state in $^{114}$Sn which is certainly not a member of the QOC quintuplet. Based on this discussion, we proposse the tenatively assigned $2^-$ state at 3396.6(2)\,keV and the $5^-$ state at 3497.9(2)\,keV as possible members of the quintuplet in $^{112}$Sn. For $^{114}$Sn, only the $3^-$ candidate at 3397.3(2)\,keV could be identified based on its $\gamma$-decay to the octupole vibrational $3^-_1$ state. We note, however, that the possible $5^-$ candidate at 3610.2(4)\,keV has only been weakly excited in our experiment. Assuming $B(E2;5^- \rightarrow 3^-_1) = B(E2;2^+_1 \rightarrow 0^+_1)$ leads to an estimated $I_{\gamma}$ of about 40\,$\%$ for $E_{\gamma} = 1335.7$\,keV. A strongly excited $3^-$ state at 3524.4(2)\,keV with a $\gamma$ transition of $E_{\gamma} = 1337.0(2)$\,keV prevented the detection of the $\gamma$-decay branch to the $3^-_1$ in the present $p\gamma$-coincidence data. No indications were found in our $\gamma\gamma$-coincidence data when applying a gate onto the $\gamma$-decay of the $3^-_1$ state. All other candidates named were cross-checked using the aforementioned $\gamma\gamma$-coincidence data.

\section{conclusion}

We have performed two inelastic proton-scattering experiments at the Institute for Nuclear Physics of the University of Cologne to study excited states in the lightest stable tin isotopes $^{112,114}$Sn. Level lifetimes and $\gamma$-decay branching ratios were determined using the combined spectroscopy setup SONIC@HORUS to acquire $p\gamma$- and $\gamma\gamma$-coincidence data.
\\

In this publication, we have studied and identified the low-spin members of the proton 2p-2h intruder configuration in $^{112}$Sn and $^{114}$Sn. With respect to their $E2$ transitions strengths, these states are more similar to corresponding states in the 0p-4h Pd nuclei than to the 4p-0h states in the Xe nuclei. Our observations are supported by $sd$ IBM-2 mixing calculcations we performed for $^{114}$Sn. Systematic calculations along these lines will further help to understand shape coexistence and isospin symmetry in the vicinity of the Sn isotopes. Especially, measuring the lifetimes of the $4^+_{\mathrm{intr.}}$ and $6^+_{\mathrm{intr.}}$ states in $^{116}$Sn would be instructive. Then, a stringent comparison of the mixing between and the evolution of the two different configurations, which we sketched for $^{114}$Sn, would be possible.
\\

Since the two-phonon states, if present at all, already mix with the intruder configuration, the identification of possible three-phonon quadrupole states is even less straight-forward in $^{112,114}$Sn. We have, however, identified possible candidates which decay very similar to the candidates previously proposed in $^{124}$Sn. Still, the $B(E2)$ strengths of these states strongly deviates from the simple vibrational picture. Further systematic experimental and theoretical investigations are highly desirable.
\\
 
Besides the coupling of quadrupole phonons, we studied possible members of the QOC quintuplet and identified candidates. The new $J^{\pi} = 1^-$ candidate in $^{114}$Sn fits nicely into the systematics established for the previously studied nuclei. Therefore, it has been clearly shown that our new method provides the means to study such structures in nuclei with low abundance where the amount of target material needed to study these with other methods, ${\it e.g.}$, $(n,n'\gamma)$ or $(\gamma,\gamma')$, is hardly affordable. The situation in $^{112}$Sn, however, remains unsatisfying. No clear agreement is observed between the different measurements. It is desirable to remeasure $^{112}$Sn with the NRF technique to check the efficiency and photon flux ambiguities previously encountered\,\cite{Pys04a,Pys06a}. Disturbingly, the expected $B(E2;1^- \rightarrow 3^-_1)$ has only been observed in a few nuclei up to now not including the Sn isotopes, see Ref.\,\cite{Der16a} and references therein. Possible $J^{\pi} = 1^-$ states have been observed close to the sum energy which would decay as expected from a QOC $1^-$ state. Firm spin-parity assignments and further investigations are needed. We have also reported candidates for the $2^-$, $3^-$ and $5^-$ members of the quintuplet making $^{112,114}$Sn the only two Sn nuclei where several members are identified based on their excitation energy and transition strengths. Therefore, future experiments using different experimental probes to identify the complete multiplet in the other Sn isotopes are highly desirable. Candidates have been reported in $^{116}$Sn\,\cite{Ram91a} and some negative-parity states at approximately the expected energies are already known in the other Sn isotopes\,\cite{ENSDF}. However, lifetime data is missing so far. As shown in this publication, the $(p,p'\gamma)$ DSA coincidence technique could be used to measure these lifetimes. Furthermore, the assignment of spin and parity will be possible with enhanced statistics as well as with the new and improved SONIC spectrometer based on $p\gamma$-angular correlations.

\begin{acknowledgments}
We gratefully acknowledge the help of the accelerator staff at the IKP Cologne. We want to thank F. Iachello, J.\,Jolie, C.\,Petrache, A. Schreckling, and N.\,Warr for helpful and stimulating discussions as well as A.\,Blazhev and K.-O.\,Zell for the target preparation. We also want to thank H.-W.\,Becker for his support during the RBS measurements as well as the NNDC for a consistency check of our data prior to publication. E.L. acknowledges support from US-NSF grant PHY-1404343 and NSF CAREER grant $\#$1654379. This work was supported by the Deutsche Forschungsgemeinschaft under Contract ZI 510/7-1.
\end{acknowledgments}

\bibliography{Sn_prc}

\end{document}